\documentclass[aps,prb,twocolumn,times,amsmath,amssymb,superscriptaddress,floatfix,longbibliography,footnoteinbib]{revtex4-1}

\usepackage{times}
\usepackage{float}
\usepackage{amsmath}
\usepackage{amsthm}
\usepackage{amssymb}
\usepackage{amsbsy}
\usepackage{graphicx}
\usepackage{pstricks}
\usepackage{color}
\usepackage{cancel}
\usepackage{subfigure} 
\usepackage{enumerate}
\usepackage{mathrsfs} 
\usepackage{multirow} 
\usepackage{wasysym}
\usepackage[colorlinks,bookmarks=false,citecolor=blue,linkcolor=red,urlcolor=blue]{hyperref}
\usepackage{sidecap}

\begin{document}

\title{Ordered ground states of kagome magnets with generic exchange anisotropy}

\author{Owen Benton}
\affiliation{RIKEN Center for Emergent Matter Science (CEMS), Wako, Saitama, 351-0198, Japan}
\affiliation{Max Planck Institute for the Physics of Complex Systems, N{\"o}thnitzer Str. 38,
Dresden 01187, Germany} 
%

\begin{abstract}
There is a growing family of rare-earth kagome materials with dominant
nearest-neighbor interactions and strong spin orbit coupling.
The low symmetry of these materials makes theoretical description complicated,
with six distinct  nearest-neighbor coupling parameters allowed.
In this Article, we ask what kinds of classical, ordered, ground states can be expected
to occur in these materials, assuming generic (i.e. non-fine-tuned) sets of
exchange parameters.
We use symmetry analysis to show that there are
only five distinct classical ground state phases occurring for generic
parameters.
The five phases are: 
(i) a coplanar, 2-fold degenerate, state with vanishing 
magnetization (${\sf A_1}$),
(ii) a noncoplanar, 2-fold degenerate, state with 
magnetization perpendicular to the kagome plane  (${\sf A_2}$),
(iii) a coplanar, 6-fold degenerate, state with 
magnetization lying within the kagome plane  (${\sf E}$-coplanar),
(iv) a noncoplanar, 6-fold degenerate, state 
with magnetization lying within a mirror plane of the lattice (${\sf E}$-noncoplanar$_{6}$),
(v) a noncoplanar, 12-fold degenerate, state with 
magnetization in an arbitrary direction
(${\sf E}$-noncoplanar$_{12}$).
All five are translation invariant (${\bf q}=0$) states. 
Having found the set of possible ground states, the ground state phase diagram is
obtained by comparing numerically optimized energies for each possibility as a function
of the coupling parameters.
The state ${\sf E}$-noncoplanar$_{12}$ is extremely rare, occupying  $<1\%$ of the full phase diagram,
so for practical purposes there are four main ordered states likely
to occur in anisotropic kagome magnets with dominant nearest neighbor interactions.
These results can aid in interpreting recent experiments on 
``tripod kagome''
systems R$_3$A$_2$Sb$_3$O$_{14}$, as well as materials closer to the isotropic
limit such as Cr- and Fe- jarosites.
\end{abstract}

\maketitle

\section{Introduction}
\label{sec:intro}

Frustration can come from various sources.
This is certainly true of the frustration
exhibited by many magnetic materials, which may be
generated by the geometry of the lattice \cite{Ramirez94a, harris97}, by 
competition between interactions of different kinds \cite{henley89, iqbal19} or by 
bond-dependent anisotropies \cite{Kitaev2006, rousochatzakis17}.
Sometimes, all of these sources of frustration
are present at once, making the problem of determining
a ground state both more challenging and more rich \cite{yan17, essafi17, zhu18}.

Kagome lattice rare-earth materials 
\cite{zorko10, ghosh14, sharma16, sanders16, sanders16-jmmc, Dun16a, scheie16, Dun17a, scheie18, ding18, scheie-arXiv19} provide a realization of
this scenario.
The kagome lattice [Fig. \ref{fig:kagome}] is paradigmatic of geometrical frustration
while the strong spin-orbit coupling inherent to many rare-earth
ions produces complicated anisotropic exchange interactions
with distinct, competing, contributions 
and bond-dependence.

In this Article we study a model of anisotropic exchange on the kagome lattice, including
all possible nearest neighbor interactions consistent with the lattice symmetries \cite{essafi17}.
This model has six independent coupling parameters, once one allows for the absence of
reflection symmetry in the kagome plane, as is appropriate for many materials.

Several previous works have investigated different types of allowed anisotropic nearest-neighbor interaction
on the kagome lattice \cite{essafi17, Elhajal2002, ballou03, cepas08, Messio10a, Chernyshev2014, gotze15, Essafi16a, changlani18, morita18, yang20}, 
but none has treated all possible interactions
at once, in the absence of reflection symmetry in the plane.
Thus, in some sense, these previous works can be viewed as higher-symmetry limits of the generic case studied here. 
Our goal in this work is to identify the ordered, classical, ground states which are stable over a finite
fraction of the six dimensional parameter space of the full model.
We will not address the physics at the phase boundaries between different states or limits featuring high
symmetry beyond time reversal and lattice symmetries, or cases of accidental degeneracy, although these
can be of interest.
In this sense, we are studying those ground states stable in the presence of ``generic'' exchange
anisotropy.

We find that in the full six-dimensional parameter space there are only five such
distinct ground states.
They are all translationally invariant, and may be classified by how they transform
under the $C_{3v}$ point group symmetries of the kagome lattice.
Example spin configurations for each are shown in Figs. \ref{fig:A1}-\ref{fig:Enoncoplanar12}.

%

In addition to materials with strong exchange anisotropy, our approach is also useful for understanding materials
where anisotropy is weak but nevertheless plays a key role in selecting the ground state due to the frustrated nature
of Heisenberg interactions on the kagome lattice.
%
Our results can be
viewed as illuminating the spectrum of possible ground states which can be obtained by perturbing an isotropic
kagome magnet with various allowed forms of nearest-neighbor exchange anisotropy.
This may be of use in understanding the ordered ground states of materials including the
Cr- and Fe- jarosites \cite{grohol03, nishiyama03, 
morimoto03, Matan06, yildirim06} and Cd-kapellasite \cite{okuma17}.

The remainder of this Article is organised as follows:
\begin{itemize}
\item{In Section \ref{sec:hamiltonian} we review the most general symmetry allowed nearest neighbor exchange
Hamiltonian for the kagome lattice \cite{yildirim06,essafi17}. We then analyse it in terms of the
irreducible representations of the point group $C_{3v}$.}
\item{Building on this symmetry analysis, in Section \ref{sec:gs}, we demonstrate the five forms of magnetic
order which may arise from the generic Hamiltonian.}
\item{In Section \ref{sec:pd} we use numerical calculations to calculate the ground state phase
diagram of the generic Hamiltonian, delineating the regions of parameter space covered by each of the
five ordered phases.}
\item{In Section \ref{sec:experiment} we discuss experimental results on kagome materials in the light of
our calculations.}
\item{In Section \ref{sec:conclusions} we close with a brief summary and discussion
of open directions for future work.}
\end{itemize}

\section{Hamiltonian and symmetry analysis}
\label{sec:hamiltonian}

\begin{figure}
\centering
\subfigure[]{
\includegraphics[width=0.4\columnwidth]{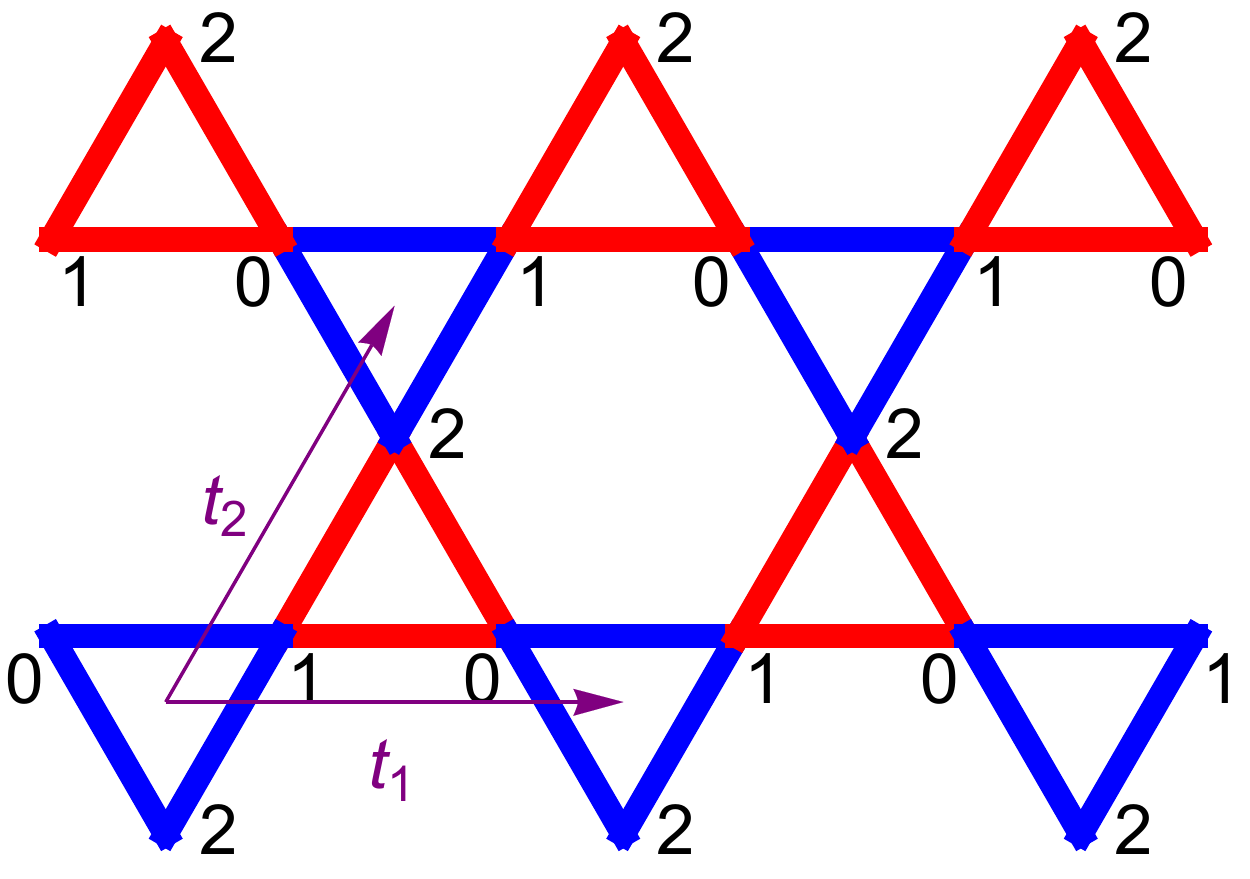}
\label{fig:kagome}
}
\subfigure[]{
\includegraphics[width=0.4\columnwidth]{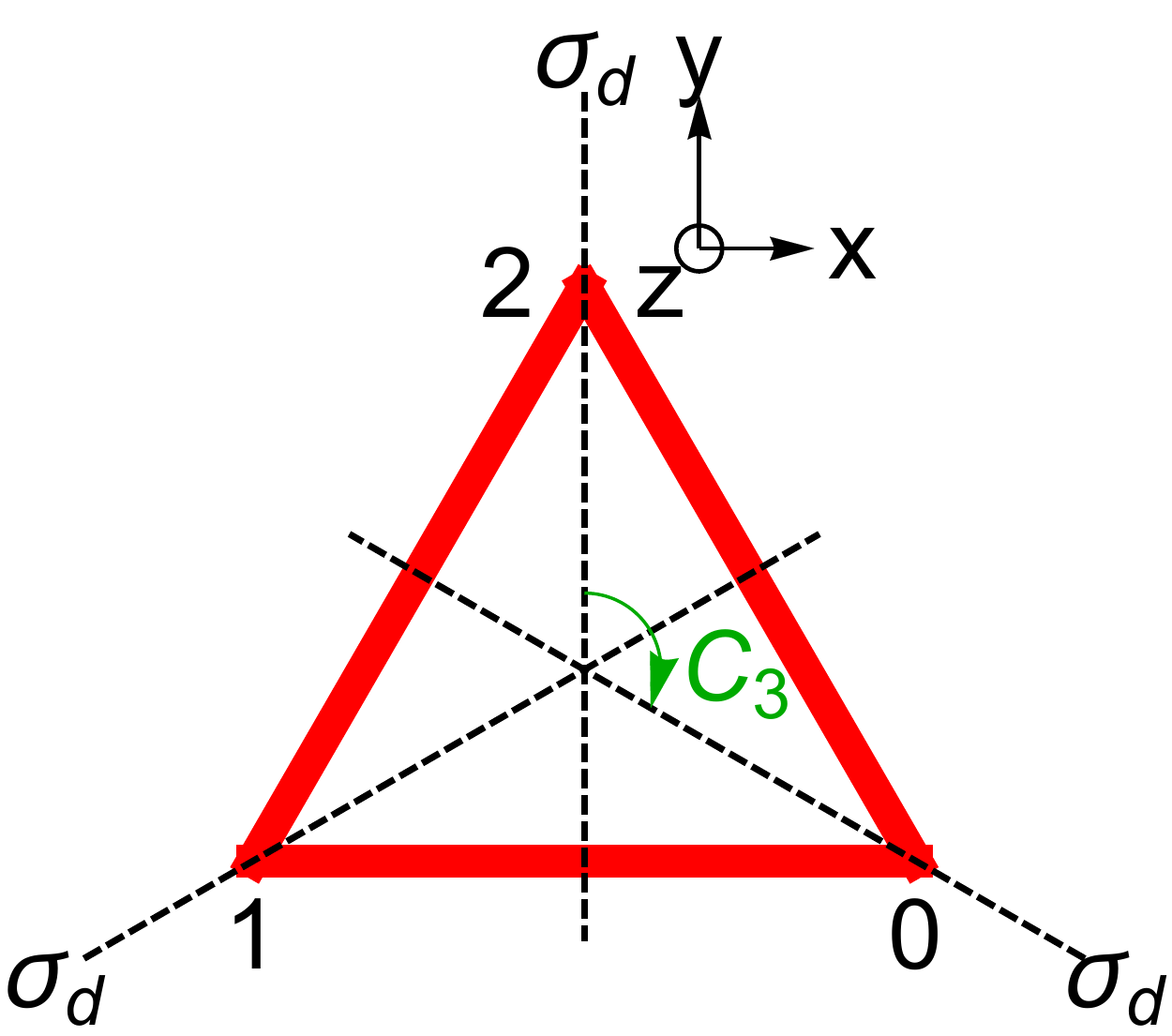}
\label{fig:triangle}
}
\caption{(a) The kagome lattice, a network of
corner sharing triangles.
The labels $0,1,2$ indicate the convention used to label the three
sublattices of the kagome lattice in this work.
 ${\bf t}_1$  and ${\bf t}_2$ are the basic translations under which the lattice is symmetric
(b) The $C_{3v}$ point group, composed of three reflection 
symmetries and a threefold rotation axis through the center of the triangle.
}
\end{figure}

We consider generalized bilinear anisotropic exchange interactions
on a kagome lattice [Fig. \ref{fig:kagome}], 
\begin{eqnarray}
\mathcal{H}=\sum_{\langle ij \rangle}
{\bf S}_i \cdot {\bf J}_{i j}  \cdot {\bf S}_j.
\label{eq:H_general}
\end{eqnarray}

We require that the interactions respect the following symmetries:
\begin{itemize}
\item{Time reversal}
\item{Lattice translations, ${\bf t}_1, {\bf t}_2$, indicated in Fig. \ref{fig:kagome}}
\item{Spatial inversion through lattice sites}
\item{$C_3$ rotations around the center of each kagome triangle [Fig. \ref{fig:triangle}]}
\item{Reflections in the mirror planes  indicated in Fig. \ref{fig:triangle}}
\end{itemize}

We do not assume any reflection symmetry in the plane of the lattice itself.

We assume that the spins ${\bf S}_i$ transform like magnetic moments, i.e. as 
axial vectors, odd under time reversal symmetry.
This will apply not only when ${\bf S}_i$ is a true magnetic moment but also when
it is a pseudospin-$1/2$ degree of freedom describing the 2-fold degenerate crystal electric 
field (CEF) ground states of a Kramers ion.
In this case the actual magnetic moment is related to the pseudospin via  the g-tensor
\begin{eqnarray}
{\bf m}_i = {\bf g}_i \cdot {\bf S}_i.
\label{eq:gtensor}
\end{eqnarray} 

An alternative case is possible in which ${\bf S}_i$ is a pseudospin describing the
low energy CEF states of a non-Kramers ion, which will generally be non-degenerate due to a lack
of protection from time reversal symmetry.
In the non-Kramers case, the pseudospin operators ${\bf S}_i$ will transform differently under
time-reversal and the discussion in this section will not apply \cite{onoda11, chen19}.

We now proceed to constrain the form of the exchange matrices ${\bf J}_{ij}$
using the symmetries listed above.
Time reversal symmetry $\mathcal{T}$
\begin{eqnarray}
\mathcal{T} {\bf S}_i = - {\bf S}_i
\end{eqnarray}
is guaranteed by the bilinear form of Eq. (\ref{eq:H_general}).

There are three spins in the unit cell, which we label
 ${\bf S}_0, {\bf S}_1, {\bf S}_2$ according
to the convention in Fig. \ref{fig:triangle}.
Translational symmetry imposes that the coupling matrices ${\bf J}_{ij}$
may only depend on which sublattices $i$ and $j$ belong to and whether the bond $ij$
is on an `up' or `down' triangle (red or blue triangles in Fig. \ref{fig:kagome}).
Inversion symmetry then guarantees that `up' and `down' triangles have the same coupling matrices.

There are thus three different coupling matrices ${\bf J}_{01},{\bf J}_{12},{\bf J}_{20}$
entering Eq. (\ref{eq:H_general}) which define the interactions between nearest
neighbour spins on each pair of sublattices.

The form of the matrices ${\bf J}_{ij}$ 
is constrained by the $C_{3v}$ point group symmetry 
at the center of each triangle [Fig. \ref{fig:triangle}], and 
was given in Refs. \onlinecite{essafi17, yildirim06}:
\begin{eqnarray}
&&{\bf J}_{01}=
\begin{pmatrix}
J_x & D_z & -D_y \\
-D_z & J_y & K \\
D_y & K & J_z \\
\end{pmatrix} 
\label{eq:J01}
\\
&&{\bf J}_{12}=\nonumber \\
&&
\begin{pmatrix}
 \frac{1}{4}(J_x+3 J_y)&  D_z +\frac{\sqrt{3}}{4}(J_x-J_y) & \frac{1}{2} (D_y+\sqrt{3} K)   \\
-D_z +\frac{\sqrt{3}}{4}(J_x-J_y)&    \frac{1}{4}( 3 J_x+ J_y) & \frac{1}{2} (\sqrt{3} D_y-K) \\
 \frac{1}{2} (-D_y+\sqrt{3} K) &  \frac{1}{2} (-\sqrt{3} D_y-K) & J_z \\
\end{pmatrix} \nonumber \\
\label{eq:J12}
\\
&&{\bf J}_{20}=\nonumber \\
&&
\begin{pmatrix}
 \frac{1}{4}(J_x+3 J_y)&  D_z +\frac{\sqrt{3}}{4}(J_y-J_x) & \frac{1}{2} (D_y-\sqrt{3} K)   \\
-D_z +\frac{\sqrt{3}}{4}(J_y-J_x)&    \frac{1}{4}( 3 J_x+ J_y) & \frac{1}{2} (-\sqrt{3} D_y-K) \\
 \frac{1}{2} (-D_y-\sqrt{3} K) &  \frac{1}{2} (\sqrt{3} D_y-K) & J_z \\
\end{pmatrix}.  \nonumber \\
\label{eq:J02}
\end{eqnarray}

There are six independent parameters in these exchange matrices:
three diagonal exchanges $J_x$, $J_y$, $J_z$, two Dzyaloshinskii-Moriya (DM)
interactions $D_y, D_z$ and one symmetric off-diagonal exchange $K$.

An additional symmetry which could, in principle, be present is
reflection
symmetry in the plane of the kagome lattice.
The presence of such a symmetry would reduce the set of allowed
exchange parameters to four, by setting $D_y=K=0$.
This case was discussed in detail in Ref. \onlinecite{essafi17}.
In this work, we will continue to assume that there is no reflection
symmetry in the kagome plane, as is appropriate
for many rare-earth kagome materials \cite{Dun17a}.
Therefore, we shall take both $D_y$ and $K$
to be nonzero.

To begin in determining the phase diagram it is helpful to rewrite
the Hamiltonian in terms of objects ${\bf m}_{\gamma, k}$ transforming according to the 
irreducible representations (irreps) of the point group.
${\bf m}_{\gamma, k}$ are defined for each triangle of the lattice, which we index using $k$.
This approach is discussed for 
 the kagome lattice  in Ref. \onlinecite{essafi17} and
the pyrochlore lattice in Ref. \onlinecite{yan17}.

These objects can function as local order parameters
for the different kinds of 3-sublattice order which we
will encounter on the phase diagram of the anisotropic
exchange model.
They also aid in the determination of the phase diagram
itself.
The appropriate objects are defined in Ref. \onlinecite{essafi17} 
but we reintroduce them here since they are essential to our
discussion.

Firstly, there is one scalar object transforming according
to the trivial $A_1$ representation of $C_{3v}$.
A nonzero average value of this field breaks none of the
point-group symmetries, only breaking time-reversal symmetry.
\begin{eqnarray}
&&m_{{\sf A_1}, k }= \frac{1}{3}\left(\frac{1}{2} S^x_{0,k} + \frac{\sqrt{3}}{2} S^y_{0,k} 
+\frac{1}{2} S^x_{1,k}   -\frac{\sqrt{3}}{2} S^y_{1,k}   -S^x_{2,k}   \right) \nonumber \\
\label{eq:ma1}
\end{eqnarray}
Here $S^{\alpha}_{i,k} $ is the $\alpha$ component of the spin belonging to sublattice $i$
and triangle $k$.

There are then two  linearly independent scalars, which transform
according to the $A_2$ representation.
A nonzero average value of these fields breaks time reversal symmetry and
all three mirror symmetries of $C_{3v}$ but preserves the 3-fold rotational
symmetry.
\begin{eqnarray}
&&m_{{\sf A_2 a}, k}=\frac{1}{3} \left( S^z_{0,k} + S^z_{1,k} +S^z_{2,k} \right) 
\label{eq:ma2a}
\\
&&m_{{\sf A_2 b}, k}=\frac{1}{3}\left(-\frac{\sqrt{3}}{2} S^x_{0,k} + \frac{1}{2} S^y_{0,k} 
+\frac{\sqrt{3}}{2} S^x_{1,k}  + \frac{1}{2}  S^y_{1,k}  -S^y_{2,k} \right)  \nonumber \\
\label{eq:ma2b}
\end{eqnarray}

Finally, there are three two-component vectors, transforming according
to the two dimensional E-irrep of  $C_{3v}$
\begin{eqnarray}
&& {\bf m}_{{\sf E a}, k}= \frac{1}{3}
\begin{pmatrix}
S^x_{0,k} + S^x_{1,k} + S^x_{2,k} \\
S^y_{0,k} + S^y_{1,k} + S^y_{2,k} 
\end{pmatrix} 
\label{eq:mea} \\
&& {\bf m}_{{\sf E b}, k}= \frac{1}{3}
\begin{pmatrix}
\frac{1}{2} S^x_{0,k} - \frac{\sqrt{3}}{2} S^y_{0,k} +
\frac{1}{2} S^x_{1,k}  +\frac{\sqrt{3}}{2} S^y_{1,k}  -S^x_{2,k} \\
-\frac{\sqrt{3}}{2} S^x_{0,k} - \frac{1}{2} S^y_{0,k} 
+\frac{\sqrt{3}}{2} S^x_{1,k}  - \frac{1}{2}  S^y_{1,k}  +S^y_{2,k}
\end{pmatrix}
\label{eq:meb} \nonumber \\
\\
&& {\bf m}_{{\sf E c}, k}= \frac{1}{3}
\begin{pmatrix}
\sqrt{\frac{3}{2}} \left( S^z_{0,k} -S^z_{1,k} \right) \\
\sqrt{2} \left( -\frac{1}{2} S^z_{0,k} - \frac{1}{2}S^z_{1,k}+ S^z_{2,k} \right)
\end{pmatrix}
\label{eq:mec}
\end{eqnarray}

In terms of these objects the Hamiltonian may be written
\begin{eqnarray}
&&\mathcal{H}
= \frac{3}{2}\sum_{k}
\Bigg[
\lambda_{\sf A_1} m_{{\sf A_1},k}^2 \nonumber \\
&&\qquad
+
\left( m_{{\sf A_2 a},k}, m_{{\sf A_2 b},k} \right)
\begin{pmatrix}
\lambda_{\sf A_2, aa} & \frac{\lambda_{\sf A_2, ab}}{2} \\
\frac{\lambda_{\sf A_2, ab}}{2} & \lambda_{ \sf A_2, bb}
\end{pmatrix}
\begin{pmatrix}
m_{{\sf A_2 a},k} \\
m_{{\sf A_2 b},k}
\end{pmatrix} \nonumber \\
&&\qquad  +
\left( {\bf m}_{{\sf E a},k}, {\bf m}_{{\sf E b},k}, {\bf m}_{{\sf E c},k} \right)
\begin{pmatrix}
\lambda_{\sf E, aa} & \frac{\lambda_{\sf E, ab}}{2} & \frac{\lambda_{\sf E, ac}}{2} \\
\frac{\lambda_{\sf E, ab}}{2} &\lambda_{\sf E, bb} & \frac{\lambda_{\sf E, bc}}{2} \\
 \frac{\lambda_{\sf E, ac}}{2}  &  \frac{\lambda_{\sf E, bc}}{2} & \lambda_{\sf E, cc} 
\end{pmatrix}
\begin{pmatrix}
 {\bf m}_{{\sf E a},k}\\
 {\bf m}_{{\sf E b},k}\\
 {\bf m}_{{\sf E c},k}
\end{pmatrix}
\Bigg] \nonumber \\
&&  =\frac{3}{2} \sum_{k}
\Bigg[
\lambda_{\sf A_1} m_{{\sf A_1},k}^2
+
\left( m_{{\sf A_2 a},k}, m_{{\sf A_2 b},k} \right)
\Lambda_{\sf A_2}
\begin{pmatrix}
m_{{\sf A_2 a},k} \\
m_{{\sf A_2 b},k}
\end{pmatrix}
\nonumber \\
&&\qquad
+ \left( {\bf m}_{{\sf E a},k}, {\bf m}_{{\sf E b},k}, {\bf m}_{{\sf E c},k} \right)
\Lambda_{\sf E}
\begin{pmatrix}
 {\bf m}_{{\sf E a},k}\\
 {\bf m}_{{\sf E b},k}\\
 {\bf m}_{{\sf E c},k}
\end{pmatrix} \Bigg]
\label{eq:Hm}
\end{eqnarray}
where $k$ indexes the triangles of the lattice and
the final term in Eq. (\ref{eq:Hm}) should be interpreted as a sum of 
9 scalar products between the vectors ${\bf m}_{{\sf E i},k}$.
The coefficients $\lambda_{\gamma}$ are:
\begin{eqnarray}
&&\lambda_{\sf A_1}=\frac{1}{2} \left( -2 \sqrt{3} D_z + J_x -3 J_y \right) 
\label{eq:lambdaA1}
\end{eqnarray}
\begin{eqnarray}
&&\lambda_{\sf A_2, aa}=2 J_z
\label{eq:lambdaA2aa}
\\
&&\lambda_{\sf A_2, bb}= \frac{1}{2} \left( -2 \sqrt{3} D_z - 3 J_x + J_y \right) 
\label{eq:lambdaA2bb}
\\
&&\lambda_{\sf A_2, ab}=2 \left(  \sqrt{3} D_y + K \right) 
\label{eq:lambdaA2ab}
\\
&&\lambda_{\sf E, aa}=J_x + J_y 
\label{eq:lambdaEaa}\\
&&\lambda_{\sf E, bb}=\sqrt{3}  D_z - \frac{J_x}{2}-\frac{J_y}{2}
\label{eq:lambdaEbb}
\\
&&\lambda_{\sf E, cc}=-J_z 
\label{eq:lambdaEcc}
\\
&&\lambda_{\sf E, ab}=J_x - J_y 
\label{eq:lambdaEab}
\\
&&\lambda_{\sf E, ac}=\sqrt{6} D_y - \sqrt{2} K 
\label{eq:lambdaEac}
\\
&&\lambda_{\sf E, bc}=\sqrt{8} K
\label{eq:lambdaEbc}
\end{eqnarray}

It is then useful to write
Eq. (\ref{eq:Hm}) in a new basis
chosen to diagonalize the matrices $\Lambda_{\sf A_2}$ and $\Lambda_{\sf E}$.
\begin{eqnarray}
&&\mathcal{H}=
\frac{3}{2}  \sum_{k}
\bigg(
\lambda_{\sf A_1} m_{{\sf A_1}, k}^2+
\omega_{\sf A_2 0} m_{{\sf A_2 0},k}^2
+
\omega_{\sf A_2 1} m_{{\sf A_2 1},k}^2
+ \nonumber  \\
&& \qquad \qquad
\omega_{\sf E 0} {\bf m}_{{\sf E 0},k}^2
+
\omega_{\sf E 1} {\bf m}_{{\sf E 1},k}^2
+
\omega_{\sf E 2} {\bf m}_{{\sf E 2},k}^2\bigg)
\label{eq:Hdiag}
\end{eqnarray}
Here $\omega_{{\sf A_2} i} (i=0, 1)$ are the
eigenvalues of $\Lambda_{\sf A_2}$ and 
$ m_{{\sf A_2} i}$ are linear combinations of
$ m_{\sf A_2 a}$ and $ m_{\sf A_2 b}$ corresponding
to the associated eigenvector of $\Lambda_{\sf A_2}$ [(Eq. \ref{eq:Hm})].
Similarly,  $\omega_{{\sf E} i}  (i=0, 1, 2)$ are the
eigenvalues of $\Lambda_{\sf E}$ and 
$ {\bf m}_{{\sf E} i}$ are linear combinations of
$ {\bf m}_{\sf E a}$, $ {\bf m}_{\sf E b}$ and  $ {\bf m}_{\sf E c}$ corresponding
to the associated eigenvector of $\Lambda_{\sf E}$.
We define, without loss of generality,
\begin{eqnarray}
\omega_{\sf A_2 0}\leq\omega_{\sf A_2 1},
\quad
\omega_{\sf E 0}\leq\omega_{\sf E 1}\leq\omega_{\sf E 2}
\label{eq:eval}
\end{eqnarray}

In this work we will treat the spins as classical vectors of
fixed length $|{\bf S}_i|=1$.
Due to this condition, the following constraint applies
to fields ${\bf m}_{\gamma}$ defined in Eqs. (\ref{eq:ma1})-(\ref{eq:mec}):
\begin{eqnarray}
\sum_{\gamma} |{{\bf m}_{\gamma, k}}|^2=1, \forall \ k
\label{eq:norm}
\end{eqnarray}
Eq. (\ref{eq:norm}) is a necessary but not sufficient condition
for the proper normalisation of the spins.

It should be emphasised that the reformulation of the problem in terms of
variables ${\bf m}_{\gamma, k}$ does not require any further assumptions beyond the nearest-neighbor
bilinear, nature of the interactions and the symmetries enumerated at the beginning of this section.

In what follows we will seek to find the classical ground states
of Eq. (\ref{eq:H_general}).

\section{What kinds of classical ground state are possible?}
\label{sec:gs}

In this section we seek to establish the possible classical
ordered phases which may occur on the ground state phase
diagram of Eq. (\ref{eq:H_general}).
Our focus is on classical ground states which are stable
over finite regions of the full 6-dimensional parameter space.
So, although there may be additional ground states which become
relevant in particular high symmetry limits of Eq. (\ref{eq:H_general}),
these are not the subject of our present discussion as they rely on fine-tuning
of parameters.

\begin{figure}
\centering
\includegraphics[width=0.7\columnwidth]{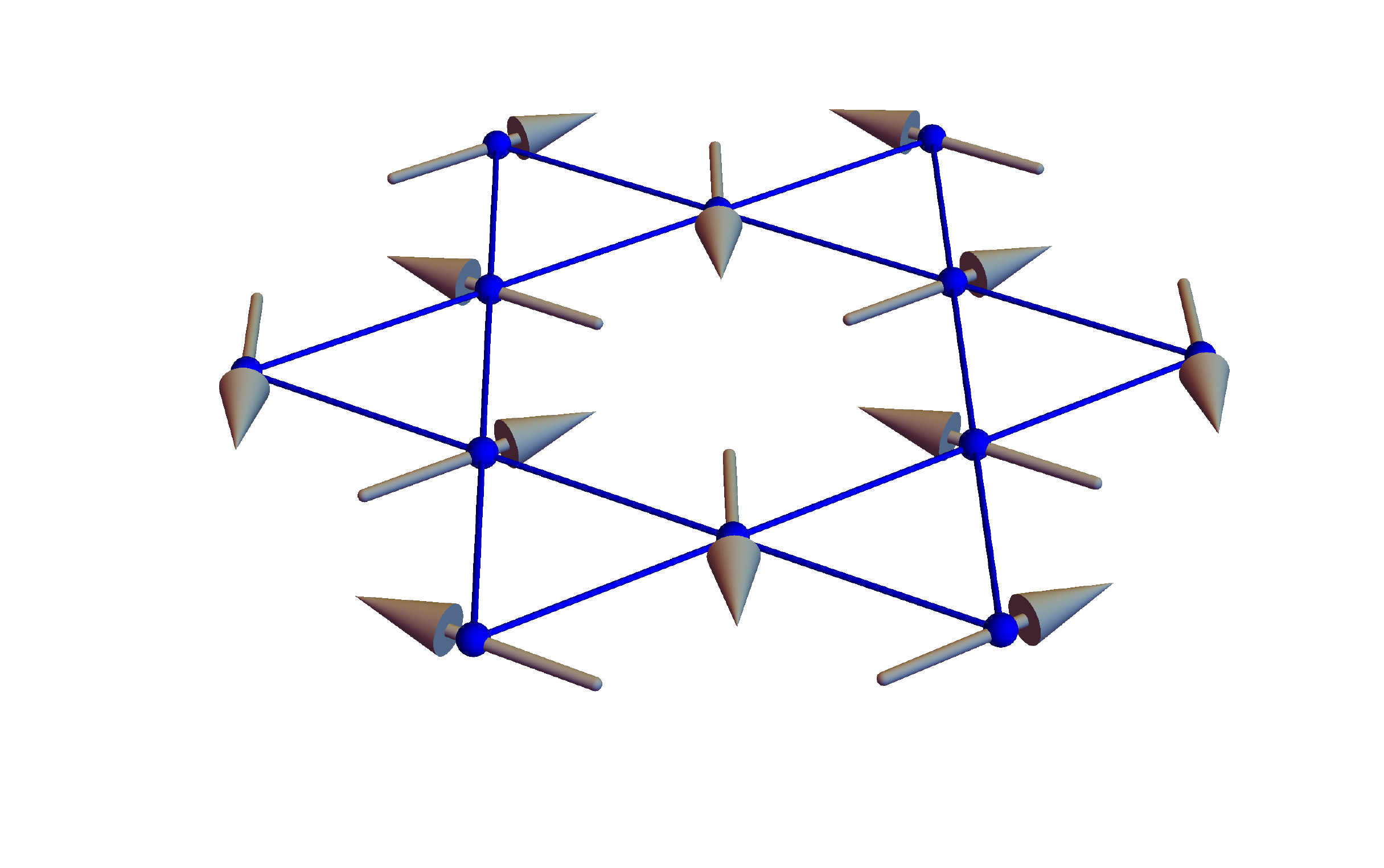}
\caption{${\sf A_1}$ ordered state, occurring as the ground state
of Eq. (\ref{eq:H_general})
when $\lambda_{\sf A_1}<\omega_{{\sf A_2} 0},\omega_{{\sf E} 0}$
[Eq. (\ref{eq:Hdiag})]. All spins lie in the kagome plane at an angle
of $\frac{2 \pi}{3}$ to one another and perpendicular to the line
joining the spin to the centers of the two neighboring triangles.
The spin configuration has vanishing total magnetization, twofold degeneracy, 
and preserves the lattice symmetries 
of Eq. (\ref{eq:H_general}) while breaking
time reversal.
}
\label{fig:A1}
\end{figure}

\begin{figure}
\centering
\includegraphics[width=0.7\columnwidth]{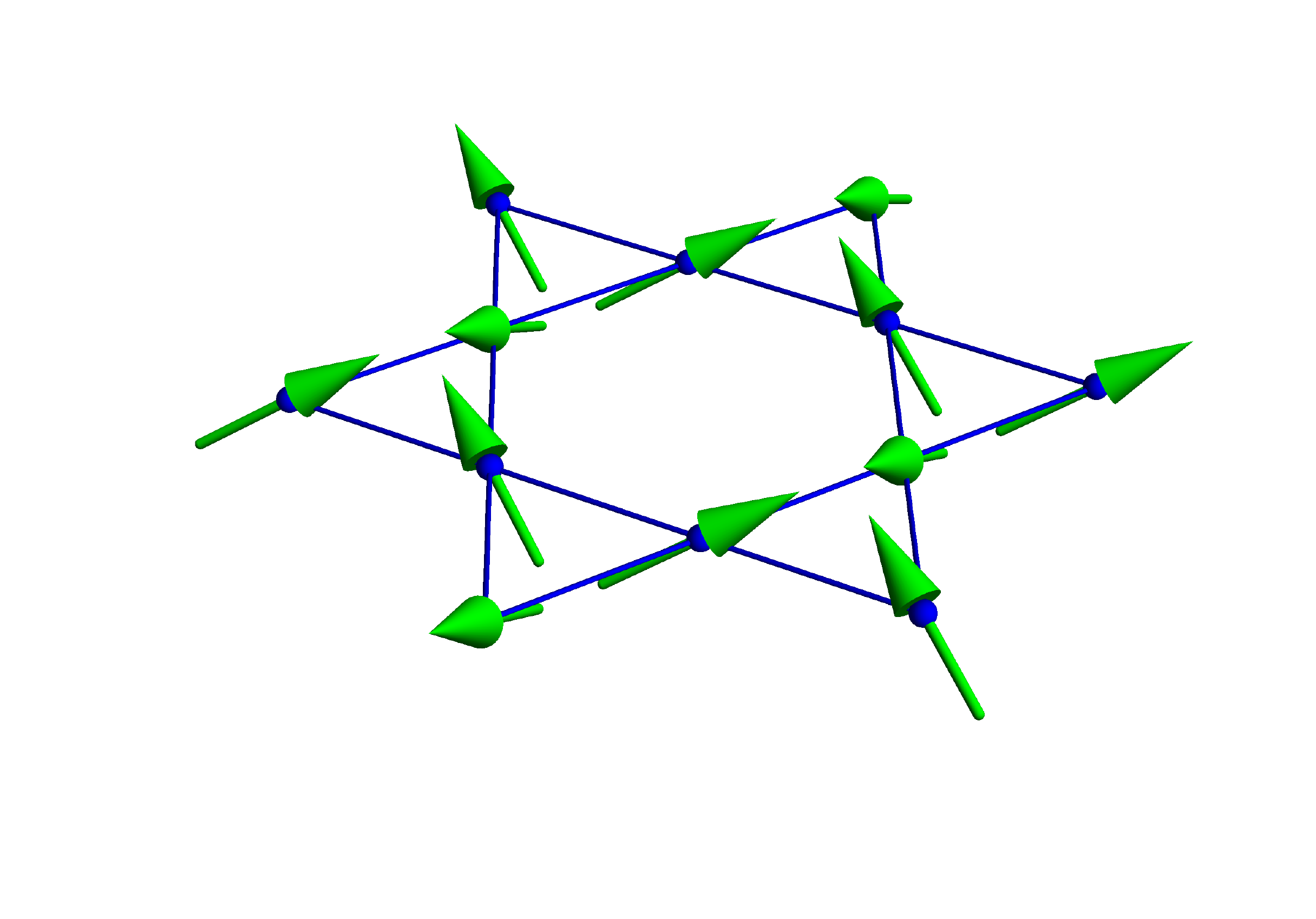}
\caption{${\sf A_2}$ ordered state, occurring as the ground state
of Eq. (\ref{eq:H_general})
when $\omega_{{\sf A_2}0}<\lambda_{\sf A_1},\omega_{{\sf E}0}$
[Eq. (\ref{eq:Hdiag})].
The spin configuration has  magnetization perpendicular to the 
kagome plane, twofold degeneracy, 
and breaks
the reflection and time reversal symmetries  
of Eq. (\ref{eq:H_general}).
}
\label{fig:A2}
\end{figure}

\begin{figure}
\centering
\includegraphics[width=0.7\columnwidth]{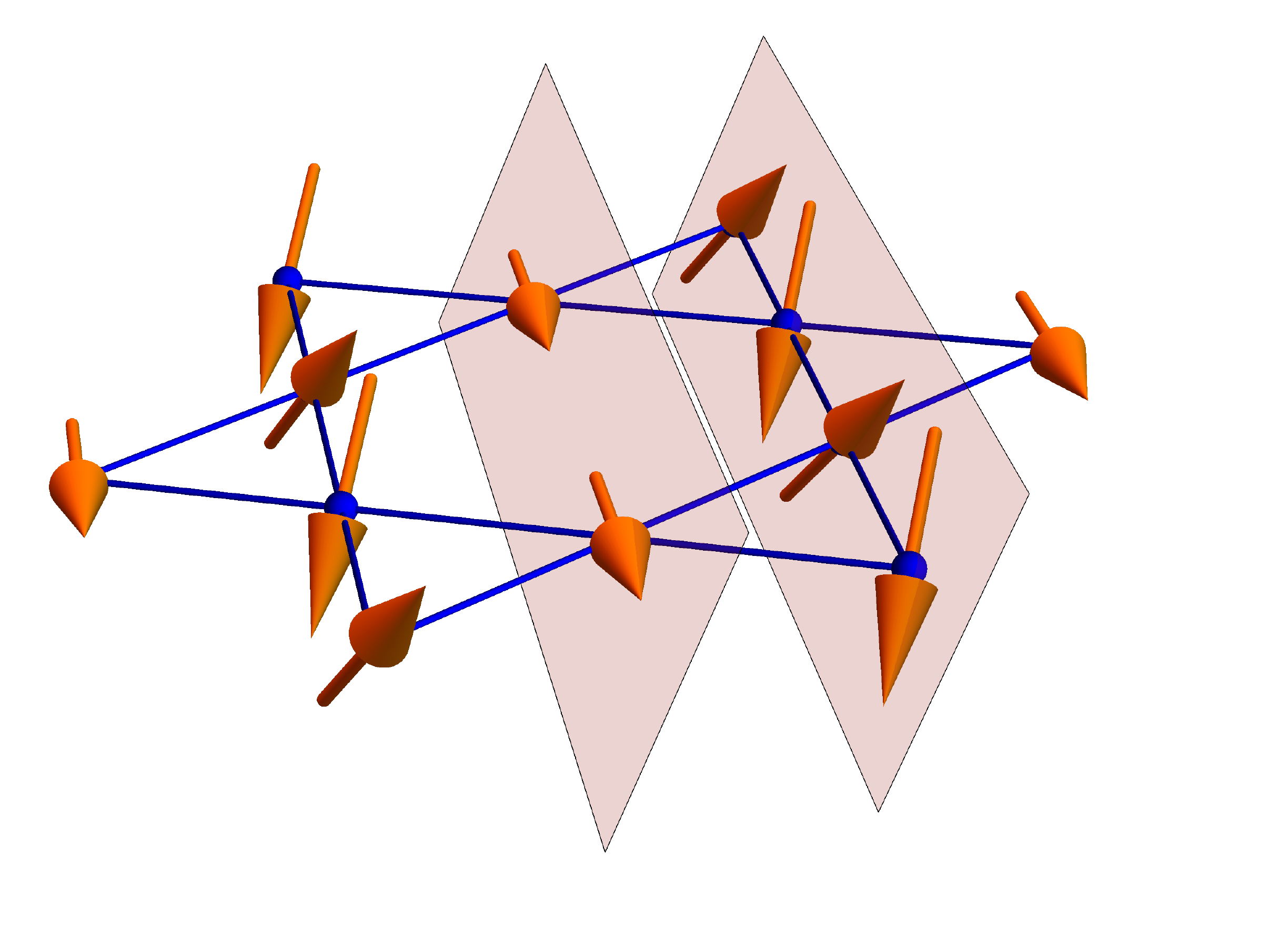}
\caption{${\sf E}$-coplanar ordered state. This occurs as a
ground state of Eq. (\ref{eq:H_general})
in part of the region where 
$\omega_{{\sf E}0}<\lambda_{\sf A_1},\omega_{{\sf A_2}0}$ [Eq. (\ref{eq:Hdiag})].
There is one spin lying in the kagome plane and two canted out
of it in such a way that the three spins remain coplanar, with the
plane of coplanarity being tilted with respect to the kagome plane. 
The plane of coplanarity is 
 indicated by the translucent red
planes.
There is a net magnetization within the kagome plane.
This state breaks time reversal and all of the point group
symmetries of the Hamiltonian, apart from a single reflection
symmetry which is preserved. 
It is sixfold degenerate.
}
\label{fig:Ecoplanar}
\end{figure}

\begin{figure}
\centering
\includegraphics[width=0.7\columnwidth]{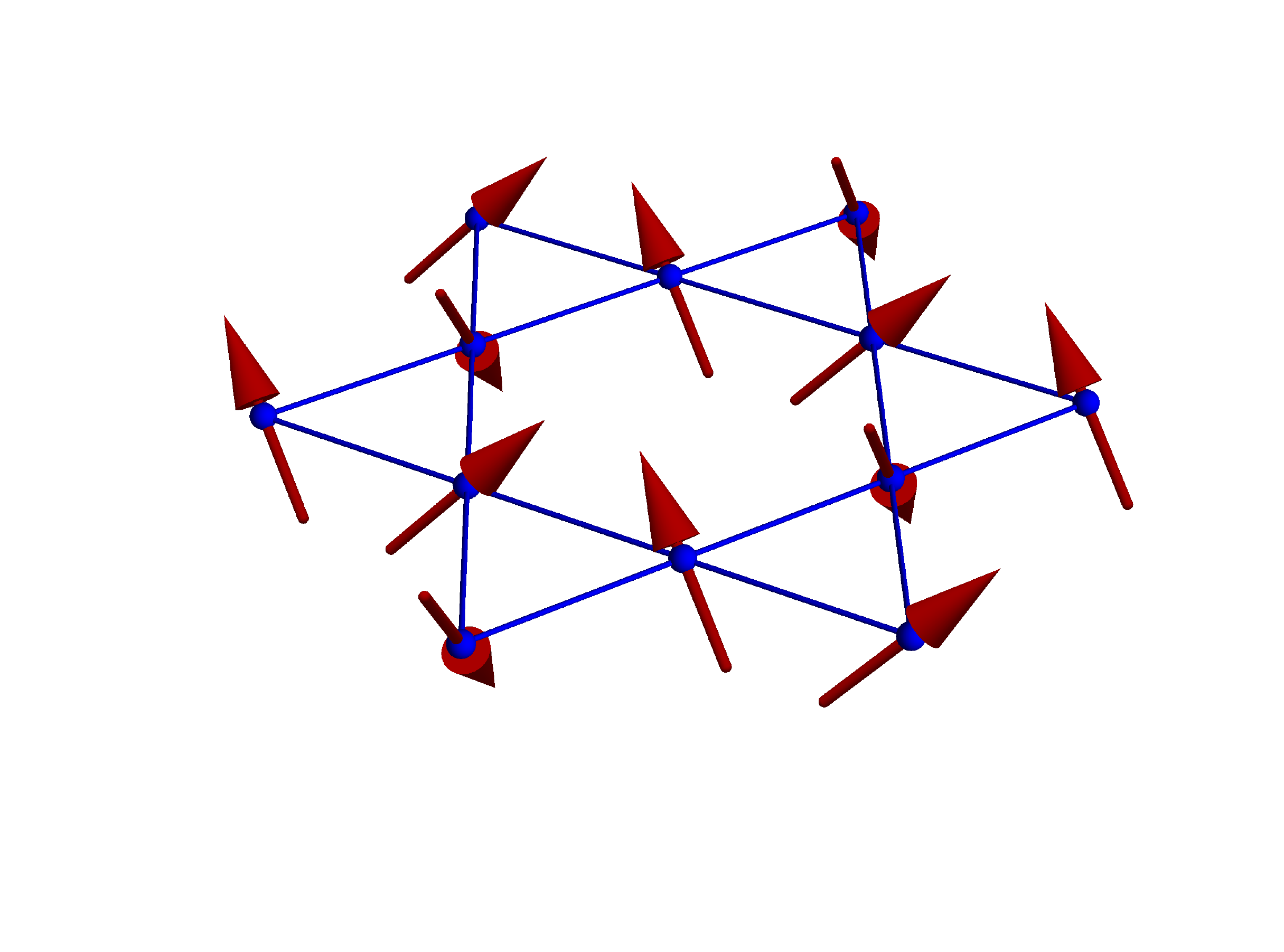}
\caption{${\sf E}$-noncoplanar$_6$ ordered state. 
This occurs as a
ground state of Eq. (\ref{eq:H_general})
in part of the region where 
$\omega_{{\sf E}0}<\lambda_{\sf A_1},\omega_{{\sf A_2}0}$ [Eq. (\ref{eq:Hdiag})].
All spins are canted out of the kagome plane
and there is a net magnetization lying within one of the mirror
planes of the lattice.
This state is non-coplanar and thus has nonzero scalar spin chirality.
This state breaks time reversal and all of the point group
symmetries of the Hamiltonian, but is symmetric under the combination
of time reversal and one  reflection
symmetry.
It is sixfold degenerate.
}
\label{fig:Enoncoplanar6}
\end{figure}

\begin{figure}
\centering
\includegraphics[width=0.7\columnwidth]{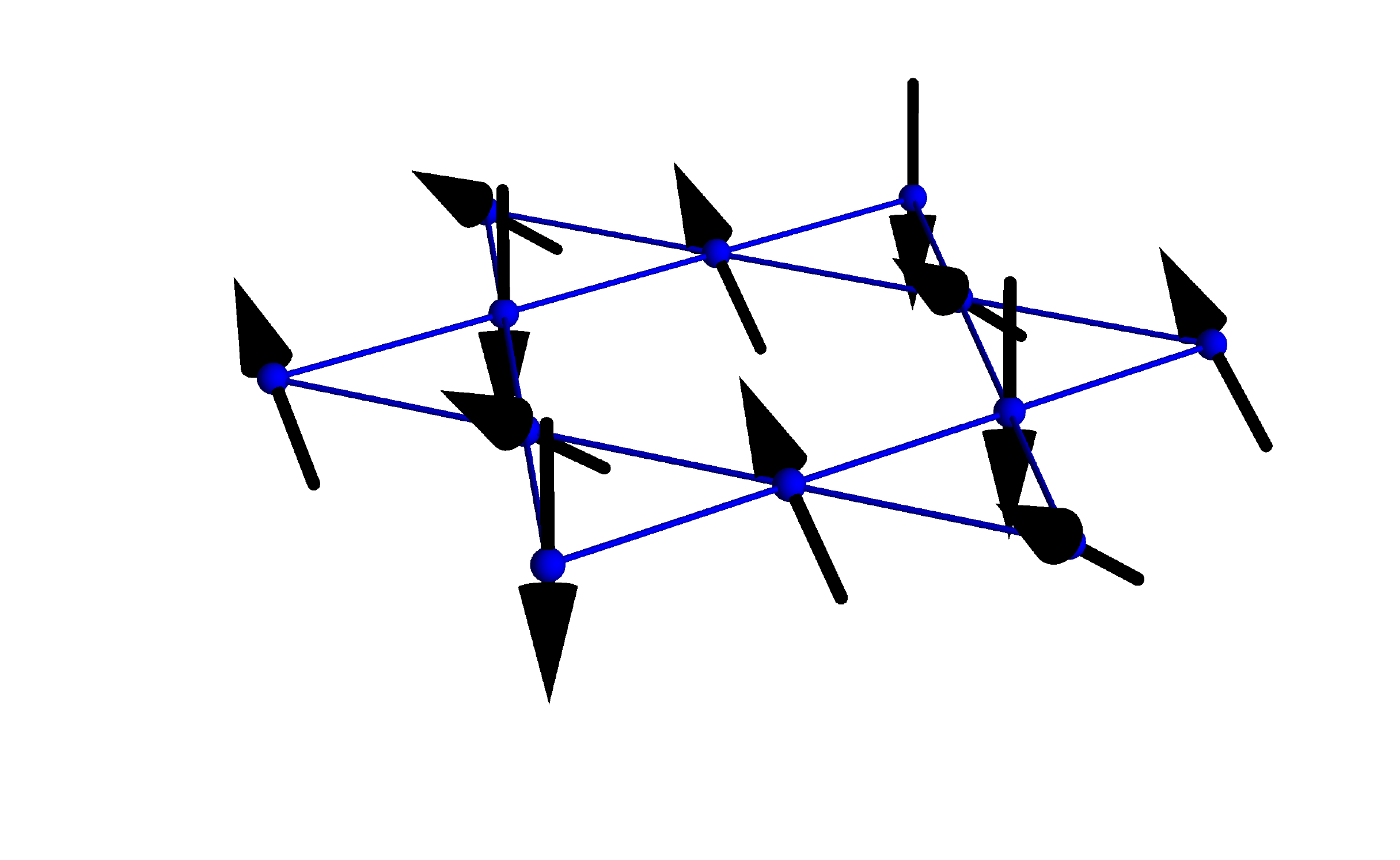}
\caption{${\sf E}$-noncoplanar$_{12}$ ordered state.
This occurs as a
ground state of Eq. (\ref{eq:H_general})
in part of the region where 
$\omega_{{\sf E}0}<\lambda_{\sf A_1},\omega_{{\sf A_2}0}$ [Eq. (\ref{eq:Hdiag})]
This state is generally non-coplanar and breaks
time reversal symmetry, all point group symmetries and all
combinations of time reversal with point group symmetries.
It is twelvefold degenerate.
This very low symmetry configuration is rare on the ground state
phase diagram, occupying $<1\%$ of the full parameter space
of the Hamiltonian [Fig. \ref{fig:piechart}].
}
\label{fig:Enoncoplanar12}
\end{figure}

Our conclusions may be summarized as follows:
\begin{enumerate}
\item{
A translation invariant (${\bf q}=0$) ground state exists for all values of exchange parameters.}
\item{If
\begin{eqnarray}
\lambda_{\sf A_1}<\omega_{{\sf A_2}0},\omega_{{\sf E}0}
\label{eq:a1condition}
\end{eqnarray}
the ground state is the antiferromagnetic state
shown in Fig. \ref{fig:A1} and discussed in Section \ref{subsec:A1}.}
\item{If
\begin{eqnarray}
\omega_{{\sf A_2}0}<\lambda_{\sf A_1},\omega_{{\sf E}0}
\label{eq:a2condition}
\end{eqnarray}
the ground state is the noncoplanar state, with magnetization perpendicular
to the plane,
shown in Fig. \ref{fig:A2} and discussed in Section \ref{subsec:A2}.}
\item{If
\begin{eqnarray}
\omega_{{\sf E}0}<\lambda_{\sf A_2},\omega_{A_1}
\label{eq:econdition}
\end{eqnarray}
the ground state may be one of three states (${\sf E}$-coplanar, 
${\sf E}$-noncoplanar$_6$,  ${\sf E}$-noncoplanar$_{12}$ 
) shown in Figs. \ref{fig:Ecoplanar}-\ref{fig:Enoncoplanar12}
and discussed in Section \ref{subsec:E}.}
\end{enumerate}

A summary of the five phases in terms of the values of local
order parameters ${{\bf m}_{\gamma}}$ [Eqs. (\ref{eq:ma1})-(\ref{eq:mec})]  is given in
Table \ref{table:ansatz}.

In what follows we will demonstrate these results.

\begin{table*}
\centering
\begin{tabular}{ | c | c |  c | c | c |  c | c | c |  c | c | c |  c | c |}
\hline
{\bf Phase} & 
$m_{\sf A_1}$
& $m_{\sf A_2a}$
& $m_{\sf A_2b}$
& $m_{\sf E a}$
& $\psi_a$
& $m_{\sf E b}$
& $\psi_b$
& $m_{\sf E c}$
& $\psi_c$
\\
\hline
${\sf A_1}$ 
& $\neq 0$
& 0 & 0 & 0 & -- & 0 &
-- &0 & --
\\
\hline
${\sf A_2}$ & $ 0$
& $\neq 0$ & $\neq 0$ & 0 & -- & 0 &
-- &0 & --
\\
\hline
${\sf E}$-coplanar & $ \neq 0$
& 0 & 0 &  $\neq 0$ & $\frac{n \pi}{3}$ & $\neq 0$ &
$\frac{n \pi}{3}$ & $ \neq 0$ & $\frac{n \pi}{3}$
\\
\hline
${\sf E}$-noncoplanar$_6$ & 0
& $ \neq 0$ & $ \neq 0$ &  $\neq 0$ & $\frac{(2n+1)  \pi}{6}$ & $\neq 0$ &
$\frac{(2n+1) \pi}{6}$ & $ \neq 0$ & $\frac{(2n+1)  \pi}{6}$ \\
\hline
${\sf E}$-noncoplanar$_{12}$ & $\neq 0$
& $ \neq 0$ & $ \neq 0$ &  $\neq 0$ & $[0, \pi)$& $\neq 0$ &
$[0, \pi)$ & $ \neq 0$ & $[0, \pi)$ \\
\hline
\end{tabular}
\caption{Description of the five possible classical ground states in
terms of the local order parameters defined for a triangle in
Eqs. (\ref{eq:ma1})-(\ref{eq:mec}).  
${\sf E}$ order parameters ${\bf m}_{\sf E_\alpha}$
are expressed in polar form, with  $m_{\sf E_{\alpha}}$
and $\psi_{\alpha}$ as defined in Eqs. (\ref{eq:Ea-angle})-(\ref{eq:Ec-angle}).
$n$ is an integer, with different choices of $n$ corresponding to different degenerate
ground states in ${\sf E}$-coplanar  and ${\sf E}$-noncoplanar$_6$ phases.
}
\label{table:ansatz}
\end{table*}

\subsection{Existence of ${\bf q}=0$ classical ground state 
for all parameter sets}
\label{subsec:q=0}

Here, for completeness, we give the proof that Eq. (\ref{eq:H_general})  possesses
a ${\bf q}=0$ classical ground state for all values of exchange parameters, 
following arguments previously given
in Refs.  \onlinecite{essafi17, yan17}.
We follow a strategy of building up the global ground state from the ground states 
of corner sharing units, as is frequently done for models on lattices with a
corner-sharing structure \cite{reimers91,essafi17,yan17,morita18,yang20}.

As we have shown above, the nearest-neighbor exchange
Hamiltonian Eq. (\ref{eq:H_general}) can be rewritten as a sum over
triangles:
\begin{eqnarray}
\mathcal{H}=\sum_{\triangle} \mathcal{H}_{\triangle}
\end{eqnarray}
with $\mathcal{H}_{\triangle}$ being the same on every triangle
of the lattice, as a consequence of inversion and translation symmetries.
This formulation makes it clear that any configuration which minimizes 
the energy of each individual triangle, also minimizes the energy of the
system as a whole.

Such a configuration may readily be obtained by minimizing the energy
on a  single ``up-pointing'' triangle  
(red triangles in Fig. \ref{fig:kagome})
and then tiling the solution over all ``up-pointing'' triangles of the lattice.
The ``up-pointing'' triangles will then all be in a ground state by construction,
and the ``down-pointing'' triangles will be too, because they have the same
exchange matrices as ``up-pointing'' triangles and the same spin orientation
on each sublattice.

This naturally results in a translation invariant (${\bf q}=0$) state, which is
guaranteed to be a ground state.
Moreover, it means that the ground state problem on the whole lattice
can be reduced to finding the ground state of three spins on a triangle.

In Sections \ref{subsec:A1}-\ref{subsec:E} we examine the various possible solutions
to this problem, that occur in different regions of parameter space.

The argument above does not rule out the existence of additional, ${\bf q}\neq0$, 
ground states, degenerate with the ${\bf q}=0$ ones. 
We regard it, however, as unlikely that such accidental degeneracies are present
over finite regions of the 6-dimensional parameter space.
Such a robust accidental degeneracy, would require a pair of states not related by any symmetry, to be
degenerate with respect to each of the six independent terms of the Hamiltonian individually,
which would seem to require a rather large 
coincidence.
The Heisenberg-Kitaev model on the kagome lattice \cite{morita18, yang20} exhibits an extended, accidental degeneracy, in the classical limit, but since that model only has two symmetry distinct
terms the required coincidence is not so large.

From now on, we assume translationally invariant ground states
built by tiling the ground states of a single triangle, and therefore drop the triangle index $k$
from the fields ${\bf m}_{\gamma,k}$ and spins ${\bf S}_{i,k}$.

We can use the solutions of the single triangle problem to check the validity
of the assumption that there are only ${\bf q}=0$ ground states.
We do this by checking whether two distinct single triangle ground states can be placed on neighboring 
triangles without causing an inconsistency at the shared site.
If they cannot, then only ${\bf q}=0$ ground states are possible.
This is explicitly checked for each single triangle ground state below, and in each case we find
that  ${\bf q}\neq0$ states are only possible with fine tuning.

\subsection{$\sf A_1$ order}
\label{subsec:A1}

We first consider the parameter regime defined by inequality (\ref{eq:a1condition})
where $\lambda_{\sf A_1}$ is the lowest coefficient in Eq. (\ref{eq:Hdiag}).

We can use Eq. (\ref{eq:norm}) to write:
\begin{eqnarray}
m_{A_1}^2=1-\sum_{\gamma\neq {\sf A_1}} m_{\gamma}^2
\label{eq:A1replace}
\end{eqnarray}
and so eliminate
$m_{\sf A_1}$ from the Hamiltonian [Eq. (\ref{eq:Hdiag})]:
\begin{eqnarray}
&&\mathcal{H}=
\frac{3}{2} \sum_{\triangle}
\bigg(
\lambda_{\sf A_1}+
(\omega_{\sf A_2 0}-\lambda_{\sf A_1}) m_{\sf A_2 0}^2
+
(\omega_{\sf A_2 1}-\lambda_{\sf A_1})  m_{\sf A_2 1}^2 
 \nonumber \\
&& \qquad
+
(\omega_{\sf E 0}-\lambda_{\sf A_1})  {\bf m}_{\sf E 0}^2
+
(\omega_{\sf E 1}-\lambda_{\sf A_1}) {\bf m}_{\sf E 1}^2
 \nonumber \\
&& \qquad
+
(\omega_{\sf E 2}-\lambda_{\sf A_1}) {\bf m}_{\sf E 2}^2\bigg)
\label{eq:Hdiag-a1}
\end{eqnarray}

All the remaining fields $m_{\sf A_2 i}, {\bf m}_{\sf E i}$
now appear as quadratic forms with positive coefficients,
due to inequalities (\ref{eq:eval})
and  (\ref{eq:a1condition}).

Therefore any spin configuration where all these fields vanish is necessarily
a ground state, for all parameter sets fulfilling the inequality (\ref{eq:a1condition}).
There are exactly two such configurations, related to each other by time
reversal symmetry:
\begin{eqnarray}
&&{\bf  S}_0=\pm\left( \frac{1}{2}, \frac{\sqrt{3}}{2}, 0 \right), \nonumber \\
&&{\bf  S}_1=\pm\left( \frac{1}{2}, -\frac{\sqrt{3}}{2}, 0 \right),  \nonumber \\
&&{\bf  S}_2=\pm\left(-1, 0, 0 \right)
\label{eq:A1config}
\end{eqnarray}
These are the ground state spin configurations of the ${\sf A_1}$ phase.
The only remaining degree of freedom on a triangle is the choice of the $+$ or $-$
sign in Eq. (\ref{eq:A1config}).
Once this sign is chosen for one triangle, consistency at the shared spin 
requires that the same sign is chosen on the neighboring triangles.
Propagating this throughout the lattice we see that only ${\bf q}=0$ tilings are possible.

This phase preserves all lattice symmetries of the original Hamiltonian but breaks
time reversal symmetry.
One of the ground states is illustrated in Fig. \ref{fig:A1}.

\subsection{$\sf A_2$ order}
\label{subsec:A2}

Next we consider parameter sets falling in the regime described by inequality
(\ref{eq:a2condition}), such that $\omega_{{\sf A_2} 0}$ is the lowest
coefficient in Eq. (\ref{eq:Hdiag}).

Under these conditions we can use Eq. (\ref{eq:norm}) to remove
$m_{\sf A_2 0}$ from the Hamiltonian  [Eq. (\ref{eq:Hdiag})] in a similar manner 
to the analysis in Section \ref{subsec:A1}.
By this means one can show that the ground states for parameter
sets obeying the inequality $(\ref{eq:a2condition})$ are of the form
\begin{eqnarray}
&&{\bf S}_0= \pm \left( -\frac{\sqrt{3}}{2} \cos(\eta), \frac{1}{2}\cos(\eta), -\sin(\eta)\right) \nonumber \\
&&{\bf S}_1=\pm\left( \frac{\sqrt{3}}{2} \cos(\eta), \frac{1}{2}\cos(\eta), -\sin(\eta)\right)  \nonumber \\
&&{\bf S}_2=\pm\left( 0,-\cos(\eta), -\sin(\eta)\right).
\label{eq:A2config}
\end{eqnarray}
With the out-of-plane canting angle $\eta$ being determined by the content of
the lowest eigenvector of $\Lambda_{\sf A_2}$ [Eq. (\ref{eq:Hm})].
In terms of the coupling parameters, $\eta$ obeys the relation:
\begin{eqnarray}
\tan(2\eta)=\frac{4 (\sqrt{3} D_y + K)}{2 \sqrt{3} D_z+3 J_x- J_y +4J_z}.
\label{eq:a2canting}
\end{eqnarray}

With $\eta$ fixed by Eq. (\ref{eq:a2canting}),
the only remaining degree of freedom on a single triangle is the choice of 
sign in Eq. (\ref{eq:A2config}).
Once this sign is chosen for one triangle, consistency at the shared spin 
requires that the same sign is chosen on the neighboring triangles.
Propagating this throughout the lattice we see that only ${\bf q}=0$ tilings are possible.

The ${\sf A}_2$ configurations have nonzero scalar chirality on the triangle:
\begin{eqnarray}
\chi=({\bf S}_0 \times {\bf S}_1)\cdot{\bf S}_2=\pm\frac{3 \sqrt{3}}{2} \cos(\eta)^2 \sin(\eta)
\end{eqnarray}

This phase breaks the reflection and
time reversal symmetry of $\mathcal{H}$ but preserves the $C_3$ rotational symmetry.
An example ground state in this phase is illustrated in Fig.~\ref{fig:A2}.

\subsection{$\sf E$ orders}
\label{subsec:E}

We then come to the case
\begin{eqnarray}
\omega_{\sf E 0} < \lambda_{\sf A_1}, \omega_{\sf A_2 0}.
\label{eq:Econdition}
\end{eqnarray}

Applying the same type of arguments as  in Sections \ref{subsec:A1}-\ref{subsec:A2}, we might
expect to find a ground state with ${m}_{\sf A_1}={m}_{\sf A_2 a}={m}_{\sf A_2 b} =0$
and with the values of ${m}_{\sf E a, b, c}$ being determined by the lowest
eigenvector of $\Lambda_{\sf E}$.
However, for typical eigenvectors of $\Lambda_{\sf E}$  this is incompatible with the spin length constraints
\begin{eqnarray}
{\bf S}_0^2={\bf S}_1^2={\bf S}_2^2=1.
\label{eq:lengths}
\end{eqnarray}

The resolution of this is that the system must mix small values of $m_{\sf A_1}, m_{\sf A_2a}, m_{\sf A_2b}$ into the ground state, so as to respect the spin length
constraints while retaining a large value of $|{\bf m}_{{\sf E} 0}|$ as favoured
by the Hamiltonian.

We can distinguish the different ways that this can happen by further consideration
of the symmetries of the problem.
Specifically, we can ask what symmetries of the Hamiltonian 
can be preserved in the presence of nonzero values of
${\bf m}_{\sf Ea}, {\bf m}_{\sf Eb}, {\bf m}_{\sf Ec}$.

There are three possibilities consistent with nonzero values of $m_{\sf E \alpha}$.
\begin{enumerate}
\item{One of the reflection symmetries of $C_{3v}$ is preserved. This corresponds 
to the $\sf E$-coplanar phase discussed below in Section \ref{subsec:Ecpl}.}
\item{The combination of one of the reflection symmetries of $C_{3v}$ 
with time reversal is preserved. This corresponds 
to the $\sf E$-noncoplanar$_6$ phase discussed below in Section \ref{subsec:Enoncpl6}.}
\item{None of the point group symmetries, nor any of their combinations
with time reversal symmetry are preserved. This corresponds 
to the $\sf E$-noncoplanar$_{12}$ phase discussed below in Section \ref{subsec:Enoncpl12}.}
\end{enumerate}

\subsubsection{$\sf E$-coplanar}
\label{subsec:Ecpl}

In the $\sf E$-coplanar phase one of the reflection symmetries of $C_{3v}$
is preserved.
For concreteness, let us suppose that the preserved  symmetry
is reflection in the $yz$ plane, i.e. the mirror plane that runs through
site 2 in Fig. \ref{fig:triangle}.
We write ${\bf m}_{\sf E a}$, ${\bf m}_{\sf E b}$, ${\bf m}_{\sf E c}$ in polar form
\begin{eqnarray}
&&{\bf m}_{\sf E a}=m_{\sf E a} 
\begin{pmatrix}
\cos(\psi_{\sf Ea}) \\
\sin(\psi_{\sf Ea})
\end{pmatrix}
\label{eq:Ea-angle}
\\
&&{\bf m}_{\sf E b}=m_{\sf E b} 
\begin{pmatrix}
\cos(\psi_{\sf Eb}) \\
\sin(\psi_{\sf Eb})
\end{pmatrix}
\label{eq:Eb-angle}
\\
&&{\bf m}_{\sf E c}=m_{\sf E c} 
\begin{pmatrix}
\cos(\psi_{\sf Ec}) \\
\sin(\psi_{\sf Ec})
\end{pmatrix}
\label{eq:Ec-angle}
\end{eqnarray}
defining the angles $\psi_{\sf Ei}$ to be lie in the interval $[0,\pi)$, and
allowing the scalars $m_{\sf Ei}$ to take either sign $\pm$.
 
Imposing preservation of reflection symmetry in the $yz$ plane constrains $\psi_{\sf E \alpha}$
\begin{eqnarray}
\psi_{\sf E\alpha}=0 \quad \forall \ \ \alpha.
\label{eq:ecoplanar-ex1}
\end{eqnarray}

More generally, if we had chosen one of the other mirror planes [Fig. \ref{fig:triangle}] 
to be preserved, we  would have $\psi_{\sf E\alpha}=\frac{n \pi}{3}, n \in \{0,1,2\}$.
If the preserved reflection plane passes through site 2 of the unit cell [see Fig. \ref{fig:kagome}]
then $n=0$, if through site 0 then $n=1$ and if through site 1 then $n=2$.

The symmetry further implies that
\begin{eqnarray}
m_{\sf A_2 a}=m_{\sf A_2 b}=0
\end{eqnarray}
but a nonzero value of $m_{\sf A_1}$ is allowed
\begin{eqnarray}
m_{\sf A_1}\neq0
\end{eqnarray}
and will be mixed into the ground state in such
a way as to satisfy the spin length constraints.
The magnitudes and relative signs of $m_{{\sf E i}}, {m}_{\sf A_1}$ are fixed by minimizing the
energy.

An example spin configuration on the three sublattices in this
phase has the form (taking $n=0$)
\begin{eqnarray}
&&{\bf S}_0=(\cos(\phi)\sin(\theta), \sin(\phi)\sin(\theta), \cos(\theta)) \nonumber  \\
&&{\bf S}_1=(\cos(\phi)\sin(\theta), -\sin(\phi)\sin(\theta), -\cos(\theta)) \nonumber \\
&&{\bf S}_2=(1, 0, 0) 
\label{eq:cpl-config}
\end{eqnarray}
where $\phi$ and $\theta$ are functions of the exchange parameters, 
which must be determined by minimizing the energy.
Degenerate spin configurations can be obtained by applying time
reversal and lattice symmetries to Eq. (\ref{eq:cpl-config}) and there
is a total degeneracy of six.

The spins are in a common plane, which is generally not the plane
of the kagome lattice.
The total magnetization of the configuration is normal to the unbroken mirror plane.
An example configuration is shown in Fig. \ref{fig:Ecoplanar}.

Minimizing the energy with respect to $\theta$ and $\phi$ gives a pair
of equations which relate the ground state canting angles to the coupling parameters.
\begin{eqnarray}
&&\frac{\partial E}{\partial \theta} =0 \implies \nonumber \\
&& \frac{1}{2} \cos(\theta) (\cos(\phi)+4\cos(\phi)^2 \sin(\theta) -\sqrt{3} \sin(\phi)) J_x + \nonumber \\
&&  \frac{1}{2} \cos(\theta) (3 \cos(\phi) + \sin(\phi) (\sqrt{3}-4\sin(\theta) \sin(\phi))) J_y + \nonumber \\
&& \sin (2\theta) J_z + (2 \cos(2\theta) \cos(\phi) -\sin(\theta)) D_y  + \nonumber \\
&& 2 \cos(\theta) (1-2\cos(\phi) \sin(\theta)) \sin(\phi) D_z  + \nonumber \\
&& (\sqrt{3} \sin(\theta) - 2 \cos(2\theta) \sin(\phi)) K=0
\label{eq:dEdtheta}\\
&&\frac{\partial E}{\partial \phi} =0 \implies \nonumber \\
&& -\frac{1}{2} \sin(\theta) (\sqrt{3} \cos(\phi) +\sin(\phi) + 2 \sin(\theta) \sin(2 \phi) ) J_x + \nonumber \\
&& \frac{1}{2} \sin(\theta) (\sqrt{3} \cos(\phi) -3\sin(\phi) - 2 \sin(\theta) \sin(2 \phi) ) J_y + \nonumber \\
&& -2 \cos(\theta) \sin(\theta) \sin(\phi) D_y + \nonumber \\
&& 2 \sin(\theta) (\cos(\phi) - \cos(2\phi) \sin(\theta)) D_z -\nonumber \\
&& 2 \cos(\theta) \cos(\phi) \sin(\theta) K =0
\label{eq:dEdphi}
\end{eqnarray}
If the angles $\theta$ and $\phi$ are measured for a given material (e.g. from refinement of Bragg
peaks) then Eqs. (\ref{eq:dEdtheta})-(\ref{eq:dEdphi}) can be used to give constraints on the coupling parameters, at least
at the level of a classical description.

Unless the angles $\phi, \theta$ are fine tuned to special values (which requires
fine tuning of exchange parameters), there is no way to place different members of the set
of 6 single-triangle ground states on neighboring triangles.
This implies that only ${\bf q}=0$ configurations are possible within this phase, for generic 
parameters.

\subsubsection{$\sf E$-noncoplanar$_6$}
\label{subsec:Enoncpl6}

In the $\sf E$-noncoplanar$_6$ phase 
the combination of
time reversal with
one of the reflection symmetries of $C_{3v}$ is preserved.
For concreteness, let us suppose the preserved  symmetry
is  the combination of time reversal with
the mirror plane that runs through
site 2 in Fig. \ref{fig:triangle}.

This symmetry constrains the angles $\psi_{\sf E \alpha}$
[Eqs. (\ref{eq:Ea-angle})-(\ref{eq:Ec-angle})], remembering that $\psi_{\sf E \alpha}$ is defined to lie in the interval $[0,\pi)$:
\begin{eqnarray}
\psi_{\sf E\alpha}=\pi/2  \quad \forall \ \ \alpha.
\end{eqnarray}
More generally, if we had chosen one of the other mirror planes 
[Fig. \ref{fig:triangle}] 
to be preserved when in combination with $\mathcal{T}$, we  would have $\psi_{\sf E\alpha}=\frac{(2 n + 1) \pi}{6}, n \in \{0,1,2\} $.
If the mirror plane preserved in combination with $\mathcal{T}$ runs through site 2 of the unit cell
[see Fig. \ref{fig:kagome}] then $n=1$, if through site 0 then $n=2$, if through site 1 then $n=0$.

The symmetry implies that
\begin{eqnarray}
m_{\sf A_1}=0
\end{eqnarray}
but nonzero values of $m_{\sf A_2 a}$ and $m_{\sf A_2 b}$
appear in the ground state as a way to satisfy the spin length constraints
\begin{eqnarray}
m_{\sf A_2}, m_{\sf A_2 b} \neq 0.
\end{eqnarray}

An example spin configuration for this phase is
\begin{eqnarray}
&&{\bf S}_0=(\cos(\nu)\sin(\mu), \sin(\nu)\sin(\mu), \cos(\mu)) \nonumber \\
&&{\bf S}_1=(-\cos(\nu)\sin(\mu), \sin(\nu)\sin(\mu), \cos(\mu)) \nonumber \\
&&{\bf S}_2=(0, \cos(\kappa), \sin(\kappa)) 
\label{eq:noncpl-config}
\end{eqnarray}
The parameters $\nu$, $\mu$ and $\kappa$ are functions of the exchange parameters and must be determined by minimizing the energy.
The ${\sf E}$-noncoplanar$_6$ configurations have nonzero scalar chirality on the triangle:
\begin{eqnarray}
&&\chi=({\bf S}_0 \times {\bf S}_1)\cdot{\bf S}_2 \nonumber \\
&&=\pm 2 \cos(\nu) \sin(\mu) (-\cos(\kappa) \cos(\mu) +\sin(\kappa) \sin(\mu) \sin(\nu))
\nonumber\\
\end{eqnarray}
The magnetization of the configuration lies within the mirror plane which is
unbroken when combined with time reversal.

Degenerate spin configurations can be obtained by applying time
reversal and lattice symmetries to Eq. (\ref{eq:noncpl-config}) and there
is a total degeneracy of six.

Minimizing the ground state energy with respect to $\nu, \mu, \kappa$ gives three
constraints relating the canting angles to the coupling parameters
\begin{multline}
\frac{\partial E}{\partial \nu}=0 \implies \\
\frac{ \sin(\mu) }{2}\left( \cos(\kappa) (3 \cos(\nu) + \sqrt{3} \sin(\nu)) + 2 \sin(\mu) \sin(2 \nu) \right) J_x  +\\ 
\frac{ \sin(\mu) }{2}\left( \cos(\kappa) (\cos(\nu) - \sqrt{3} \sin(\nu)) + 2 \sin(\mu) \sin(2 \nu) \right) J_y  + \\
\sin(\mu) (\sqrt{3} \cos(\nu) \sin(\kappa) + (2 \cos(\mu)+ \sin(\kappa)) \sin(\nu) ) D_y + \\
2 \sin(\mu) (\cos(2\nu) \sin(\mu) + \cos(\kappa) \sin(\nu) ) D_z + \\
 \sin(\mu) (2 \cos(\mu) \cos(\nu) + \sin(\kappa) (-\cos(\nu)+ \sqrt{3} \sin(\nu))) K 
=0 
\label{eq:dEdnu}\\
\end{multline}

\begin{multline}
\frac{\partial E}{\partial \mu}=0 \implies \\
\frac{-\cos(\mu)}{2} \left( 4 \cos(\nu)^2 \sin(\mu) + \cos(\kappa)(\sqrt{3} \cos(\nu) -3 \sin(\nu)) \right) J_x 
  \\
+\frac{\cos(\mu)}{2} \left( 4 \sin(\nu)^2 \sin(\mu) + \cos(\kappa)(\sqrt{3} \cos(\nu) + \sin(\nu)) \right) J_y   \\
 - 2 (\cos(\mu) + \sin(\kappa)) \sin(\mu) J_z \\
 - \big(
\cos(\nu) (2 \cos(2\mu) + \cos(\mu) \sin(\kappa))  \\
-\sqrt{3} (\cos(\kappa) \sin(\mu) + \cos(\mu)\sin(\kappa)\sin(\nu)) \big) D_y -  \\
(2 \cos(\kappa) \cos(\mu) \cos(\nu) -\sin(2\mu) \sin(2\nu)) D_z +  \bigg( \cos(\kappa) \sin(\mu) +  \\
2 \cos(2\mu) \sin(\nu) - \cos(\mu) \sin(\kappa) (\sqrt{3} \cos(\nu) + \sin(\nu))   \bigg) K  \\
=0
\label{eq:dEdmu}
\end{multline}

\begin{multline}
\frac{\partial E}{\partial \kappa}=0 \implies  \\
\frac{1}{2} \sin(\kappa) \sin(\mu) (\sqrt{3} \cos(\nu) - 3 \sin(\nu)) J_x - \\
 \frac{1}{2} \sin(\kappa) \sin(\mu) \left( \sqrt{3} \cos(\nu) + \sin(\nu) \right) J_y +
 \\
2 \cos(\kappa) \cos(\mu) J_z + \\
 (\sqrt{3} \cos(\mu) \sin(\kappa) + \cos(\kappa) \sin(\mu) (\sqrt{3} \sin(\nu) - \cos(\nu))) D_y + \\ 2 \cos(\nu) \sin(\kappa) \sin(\mu) D_z + \\
  \big( \cos(\mu) \sin(\kappa) - \cos(\kappa) \sin(\mu) (\sqrt{3} \cos(\nu) + \sin(\nu))
\big)K 
=0
\label{eq:dEdkappa}
\end{multline}

If $\nu, \mu$ and $\kappa$ are known for a system in the ${\sf E}$-noncoplanar$_6$ phase,
Eqs. (\ref{eq:dEdnu})-(\ref{eq:dEdkappa}) give three constraints on the possible coupling parameters,
within the classical description.

Different members of the set of 6 ground states cannot be placed on
neighboring triangles without causing an inconsistency, unless the angles $\mu, \nu,\kappa$
are fine tuned to special values, via fine tuning of exchange parameters.
This confirms that only ${\bf q}=0$ configurations are possible within this phase, for generic 
parameter sets.

\subsubsection{$\sf E$-noncoplanar$_{12}$}
\label{subsec:Enoncpl12}

Finally, there is the possibility that time reversal, all point group
symmetries and all combinations of the two are broken in the
ground state, leaving only translation and inversion symmetries intact.

In this case the angles $\psi_{\sf E \alpha}$ 
[Eqs. (\ref{eq:Ea-angle})-(\ref{eq:Ec-angle})] can take
arbitrary values, and symmetry does not fix any relationship between them
\begin{eqnarray}
\psi_{\sf E a} \neq \psi_{\sf E b} \neq \psi_{\sf E c}.
\end{eqnarray}
Moreover $m_{\sf A_1}, m_{\sf A_2a}, m_{\sf A_2b}$
may all be present by symmetry
\begin{eqnarray}
m_{\sf A_1}\neq0, \ \ 
m_{\sf A_2 a}\neq0, \ \ 
m_{\sf A_2 b}\neq0.
\end{eqnarray}

The spin directions of the three sites on the triangle
have no fixed relationship enforced by symmetry,
so there are 6 parameters in the ground state that can
only be determined energetically:
\begin{eqnarray}
&&{\bf S}_0=(\cos(\zeta_0)\sin(\upsilon_0), \sin(\zeta_0)\sin(\upsilon_0), \cos(\upsilon_0)) \nonumber \\
&&{\bf S}_1=(\cos(\zeta_1)\sin(\upsilon_1), \sin(\zeta_1)\sin(\upsilon_1), \cos(\upsilon_1)) \nonumber \\
&&{\bf S}_2=(\cos(\zeta_2)\sin(\upsilon_2), \sin(\zeta_2)\sin(\upsilon_2), \cos(\upsilon_2)).
\label{eq:noncpl12-config}
\end{eqnarray}

An example configuration is shown in Fig. \ref{fig:Enoncoplanar12}.
The state will generally have nonzero chirality and magnetization in an arbitrary direction.
Degenerate spin configurations can be obtained by applying time
reversal and lattice symmetries to Eq. (\ref{eq:noncpl12-config}), giving a total
degneracy of twelve - the maximum possible for a state with translation and
inversion symmetries.

As shall be shown using numerics in Section \ref{sec:pd}, this low symmetry state 
does appear on the ground state phase diagram, but only in a very small region
of parameter space.

Minimizing the energy with respect to $\zeta_i, \upsilon_i$ ($i=0,1,2$) gives a total of six equations
relating the canting angles to the coupling parameters.
\begin{eqnarray}
& &\frac{dE}{d\zeta_i}=0 \implies  \nonumber \\
&&\sum_{j \neq i} 
\begin{pmatrix}
-\sin(\zeta_i) \sin(\upsilon_i) \\ 
\cos(\zeta_i) \sin(\upsilon_i)\\
0
\end{pmatrix} \cdot {\bf J}_{ij} \cdot \begin{pmatrix}
\cos(\zeta_j) \sin(\upsilon_j)\\
\sin(\zeta_j) \sin(\upsilon_j)\\
\cos(\upsilon_j)
\end{pmatrix} =0 \nonumber \\ 
\label{eq:dEdzeta}\\
&&\frac{dE}{d\upsilon_i}=0 \implies  \nonumber \\
&&\sum_{j \neq i} 
\begin{pmatrix}
\cos(\zeta_i) \cos(\upsilon_i) \\ 
\sin(\zeta_i) \cos(\upsilon_i)\\
-\sin(\upsilon_i)
\end{pmatrix} \cdot {\bf J}_{ij} \cdot \begin{pmatrix}
\cos(\zeta_j) \sin(\upsilon_j)\\
\sin(\zeta_j) \sin(\upsilon_j)\\
\cos(\upsilon_j)
\end{pmatrix} =0 \nonumber \\
\label{eq:dEdupsilon}
\end{eqnarray}
Thus, if for a system in the ${\sf E}$-noncoplanar$_{12}$ phase, all six angles are known it should
be possible to use Eqs. (\ref{eq:dEdzeta})-(\ref{eq:dEdupsilon}) to uniquely determine the
six exchange parameters.

\begin{figure}
\centering
\includegraphics[width=0.3\textwidth]{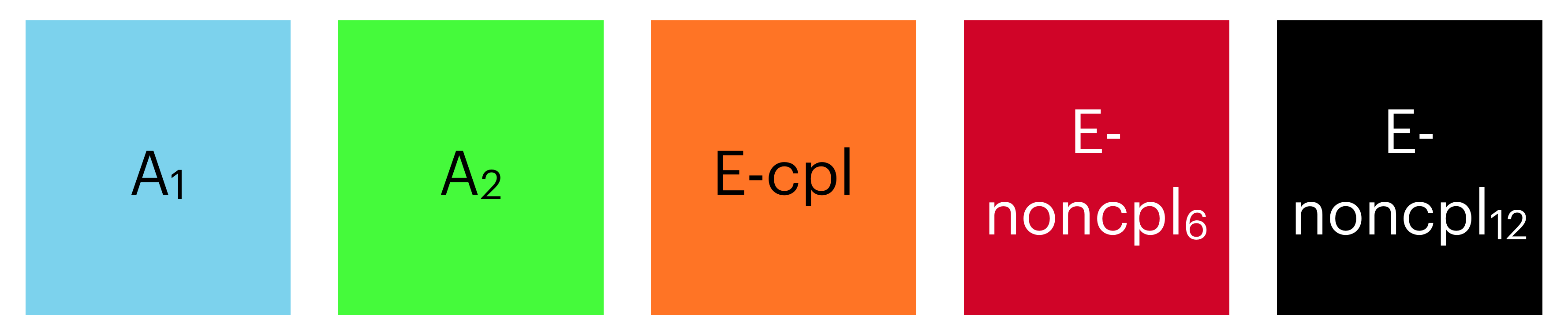}\\
\includegraphics[width=\columnwidth]{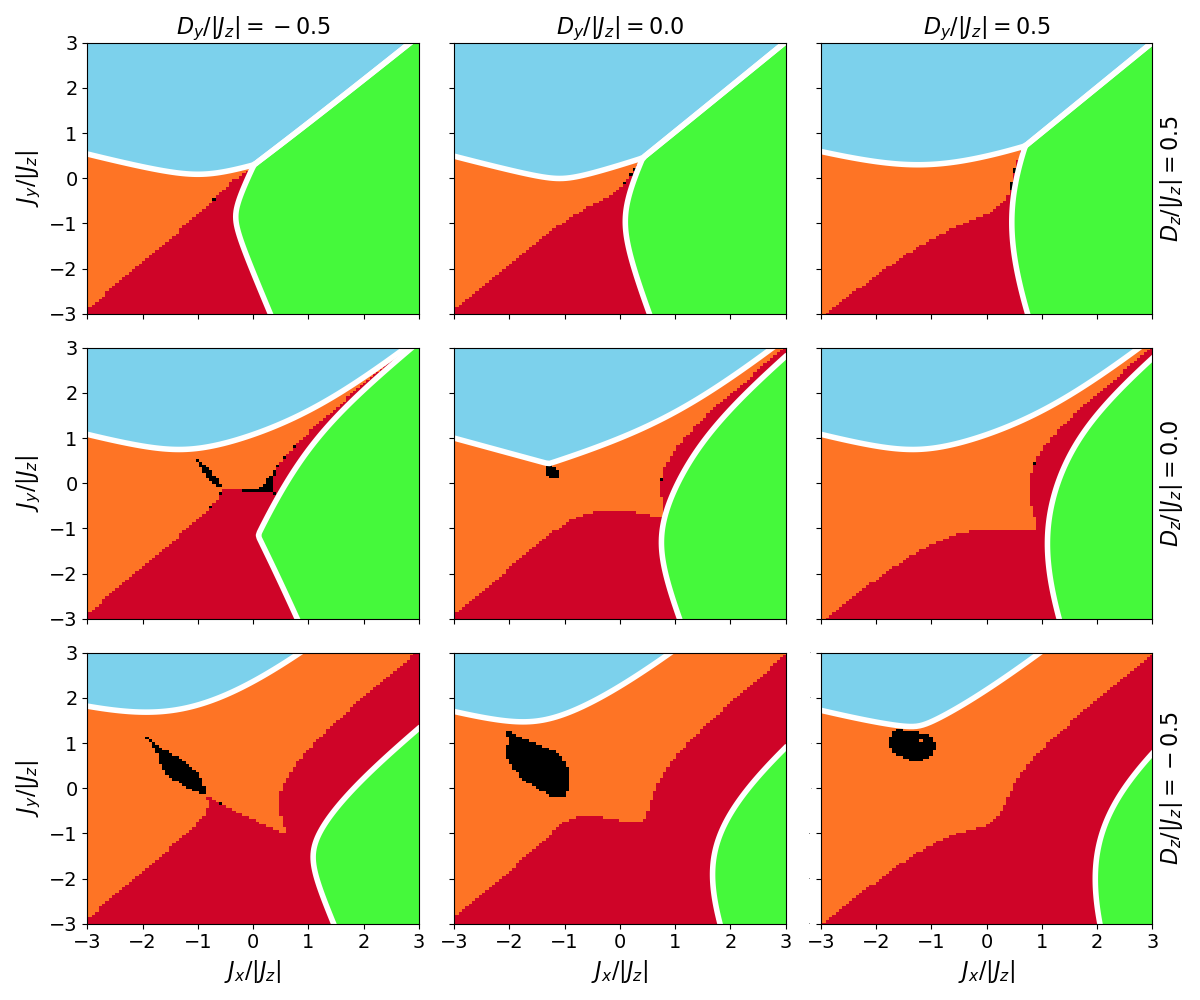}
\caption{$T=0$ phase diagram with $J_z>0$ and $K=-0.5 |J_z|$.
Each panel shows a slice of the phase diagram as a function of $J_x$ and $J_y$ for different, 
fixed, values of the DM directions $D_y$ and $D_z$, with $D_y$ increasing from left to right and
$D_z$ from bottom to top.
The phase diagram is obtained by comparing numerically optimized energies for the five phases described
in Section \ref{sec:gs}.
The numerical optimization procedure is described in Appendix \ref{app:numerics}.
The white lines show analytic calculations of the boundaries of the $\sf A_1$ and $\sf A_2$
phases, using conditions (\ref{eq:a1condition}) and (\ref{eq:a2condition}).
}
\label{fig:pd_j+_k-050}
\end{figure}

\begin{figure}
\centering
\includegraphics[width=0.3\textwidth]{color_key.pdf}\\
\includegraphics[width=\columnwidth]{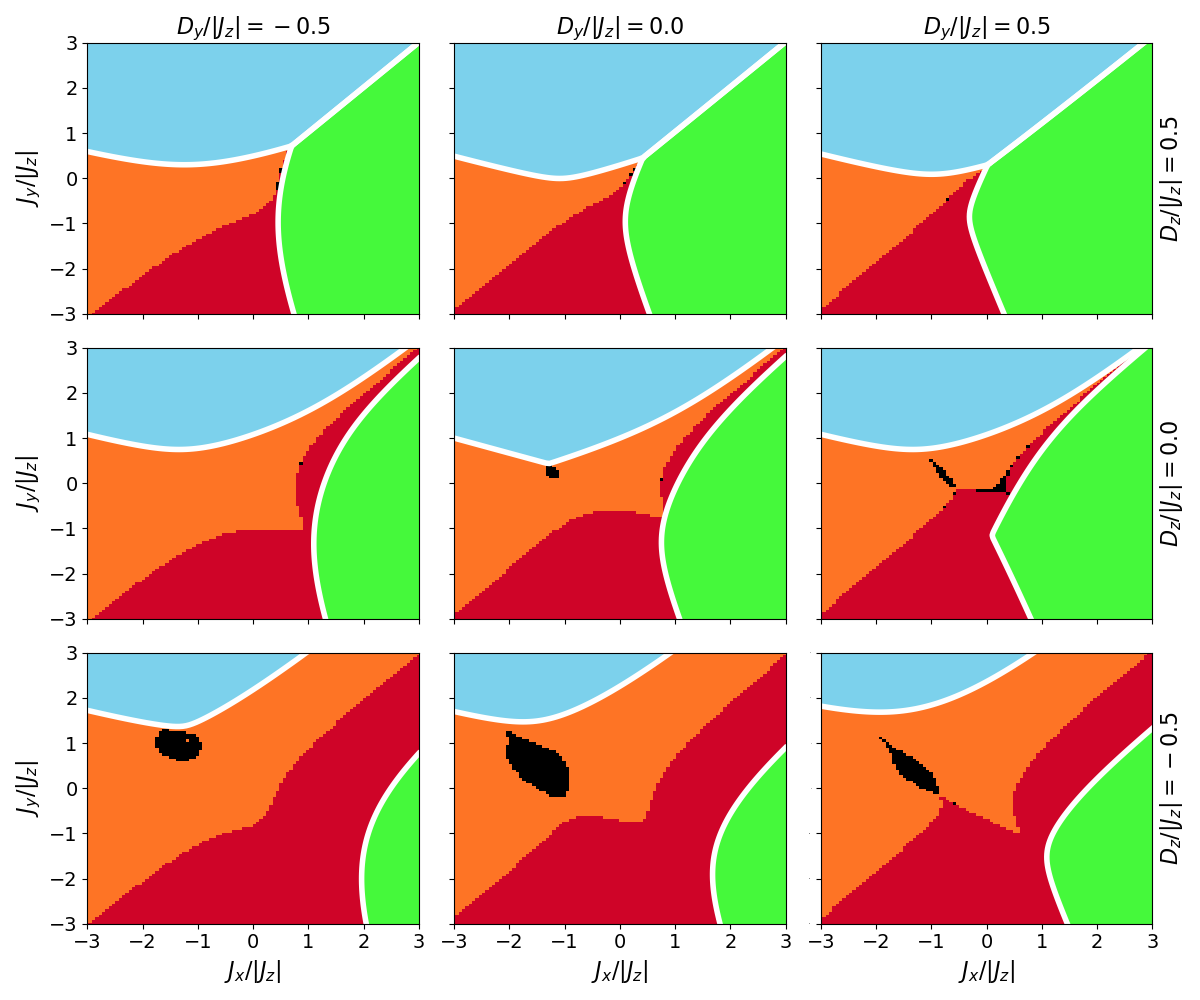}
\caption{$T=0$ phase diagram with $J_z>0$ and $K=0.5 |J_z|$.
Each panel shows a slice of the phase diagram as a function of $J_x$ and $J_y$ for different, 
fixed, values of the DM directions $D_y$ and $D_z$, with $D_y$ increasing from left to right and
$D_z$ from bottom to top.
The phase diagram is obtained by comparing numerically optimized energies for the five phases described
in Section \ref{sec:gs}.
The numerical optimization procedure is described in Appendix \ref{app:numerics}.
The white lines show analytic calculations of the boundaries of the $\sf A_1$ and $\sf A_2$
phases, using conditions (\ref{eq:a1condition}) and (\ref{eq:a2condition}).
}
\label{fig:pd_j+_k+050}
\end{figure}

\section{Phase Diagram}
\label{sec:pd}

In this section we calculate the ground state phase diagram of
Eq. (\ref{eq:H_general}) numerically, by comparing optimized
energies for the five phases described in Section \ref{sec:gs}.
The numerical optimization of the energy was done by
a combination of random search, simulated annealing and iterative minimisation \cite{sim18}.
Details of the numerics are given in Appendix \ref{app:numerics}.

\begin{figure}
\centering
\includegraphics[width=0.3\textwidth]{color_key.pdf}\\
\includegraphics[width=\columnwidth]{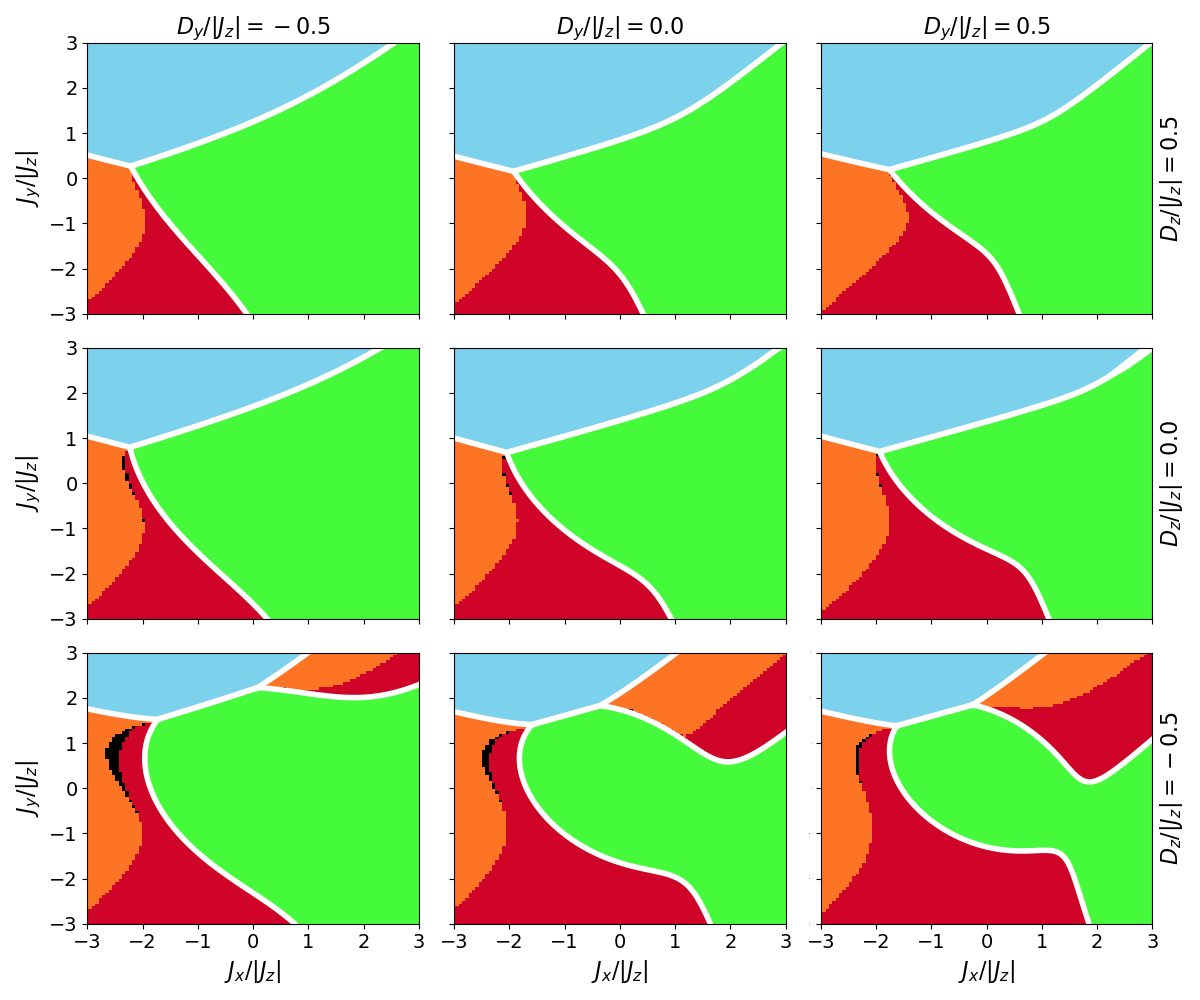}
\caption{$T=0$ phase diagram with $J_z<0$ and $K=-0.5 |J_z|$.
Each panel shows a slice of the phase diagram as a function of $J_x$ and $J_y$ for different, 
fixed, values of the DM directions $D_y$ and $D_z$, with $D_y$ increasing from left to right and
$D_z$ from bottom to top.
The phase diagram is obtained by comparing numerically optimized energies for the five phases described
in Section \ref{sec:gs}.
The numerical optimization procedure is described in Appendix \ref{app:numerics}.
The white lines show analytic calculations of the boundaries of the $\sf A_1$ and $\sf A_2$
phases, using conditions (\ref{eq:a1condition}) and (\ref{eq:a2condition}).
}
\label{fig:pd_j-_k-050}
\end{figure}

\begin{figure}
\includegraphics[width=0.3\textwidth]{color_key.pdf}\\
\includegraphics[width=\columnwidth]{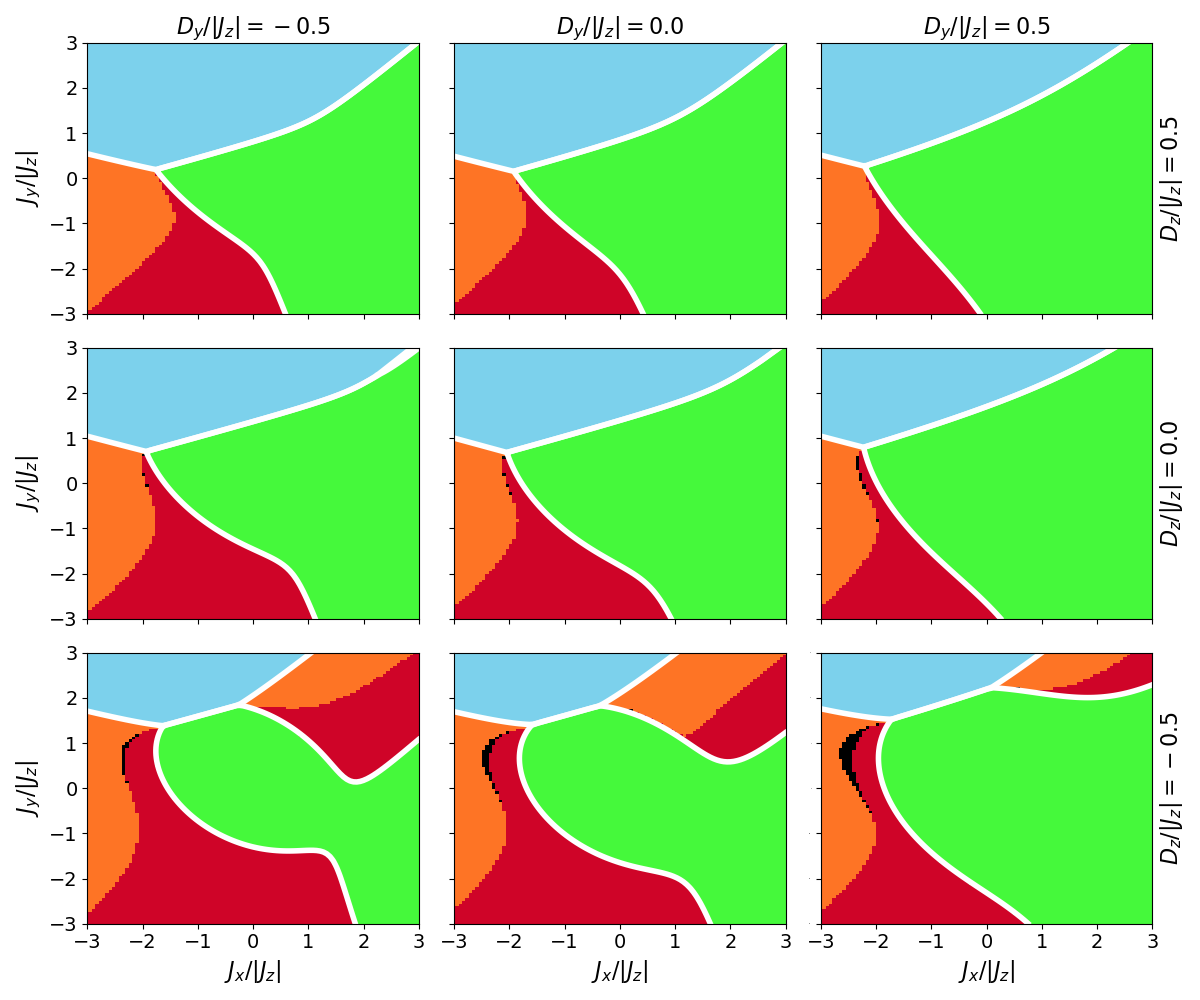}
\caption{$T=0$ phase diagram with $J_z<0$ and $K=0.5 |J_z|$.
Each panel shows a slice of the phase diagram as a function of $J_x$ and $J_y$ for different, 
fixed, values of the DM directions $D_y$ and $D_z$, with $D_y$ increasing from left to right and
$D_z$ from bottom to top.
The phase diagram is obtained by comparing numerically optimized energies for the five phases described
in Section \ref{sec:gs}.
The numerical optimization procedure is described in Appendix \ref{app:numerics}.
The white lines show analytic calculations of the boundaries of the $\sf A_1$ and $\sf A_2$
phases, using conditions (\ref{eq:a1condition}) and (\ref{eq:a2condition}).
}
\label{fig:pd_j-_k+050}
\end{figure}

Figs. \ref{fig:pd_j+_k-050}-\ref{fig:pd_j-_k+050} show slices of the phase diagram 
as a function of $J_x/|J_z|$ and  $J_y/|J_z|$ with $K/|J_z|=\{-0.5, 0.5 \}$
for both positive [Figs.  \ref{fig:pd_j+_k-050}- \ref{fig:pd_j+_k+050}]
and negative  [Figs.  \ref{fig:pd_j-_k-050}- \ref{fig:pd_j-_k+050}] $J_z$.
Each panel in a given figure corresponds to different values of DM interactions $D_y/|J_z|$
and $D_z/|J_z|$. 
$D_y/|J_z|$ increases from left to right within each figure and $D_z/|J_z|$ from bottom to top.
Taken together, Figs. \ref{fig:pd_j+_k-050}-\ref{fig:pd_j-_k+050} give a broad view of the competition between different magnetic
orders as anisotropic exchange parameters are varied.
Further phase diagrams, for a  greater range and variety of parameter sets are shown in the Supplemental Material \footnote{See the Supplemental Material for a series phase diagrams as a function of $J_x/|J_z|$ and $J_y/|J_z|$, with values of $D_y/|J_z|, D_z/|J_z|, K/|J_z|$ varying over $\{-0.75, -0.25, 0.25, 0.75\}$, for both signs of $J_z$. }.


The boundaries of the $A_1$ and $A_2$ phases can also be calculated analytically using conditions
(\ref{eq:a1condition}) and (\ref{eq:a2condition}).
These analytic boundaries are shown as white lines in Figs. \ref{fig:pd_j+_k-050}-\ref{fig:pd_j-_k+050},
and agree with the results of the numerics.
The boundaries between the different $\sf E$ phases are only calculated numerically.

One notable feature of Figs. \ref{fig:pd_j+_k-050}-\ref{fig:pd_j-_k+050} is that
the ${\sf E}$-coplanar phase is generally found bordering the ${\sf A}_1$ phase,
whereas the ${\sf E}$-noncoplanar$_6$ phase is generally found 
bordering the ${\sf A}_2$ phase.
This is natural since the ${\sf E}$-coplanar phase mixes in a finite value of the ${\sf A}_1$
order parameter and likewise the ${\sf E}$-noncoplanar$_6$ includes a finite ${\sf A}_2$
order parameter.

Another striking feature of the phase diagram is the rarity of the 
${\sf E}$-noncoplanar$_{12}$ phase.
This low-symmetry configuration occupies only small portions of
the phase diagrams in Figs. \ref{fig:pd_j+_k-050}-\ref{fig:pd_j-_k+050}, with its stability generally
being increased by a strong negative value of $D_z$.

\begin{figure}
\centering
\subfigure[]{
\includegraphics[width=\columnwidth]{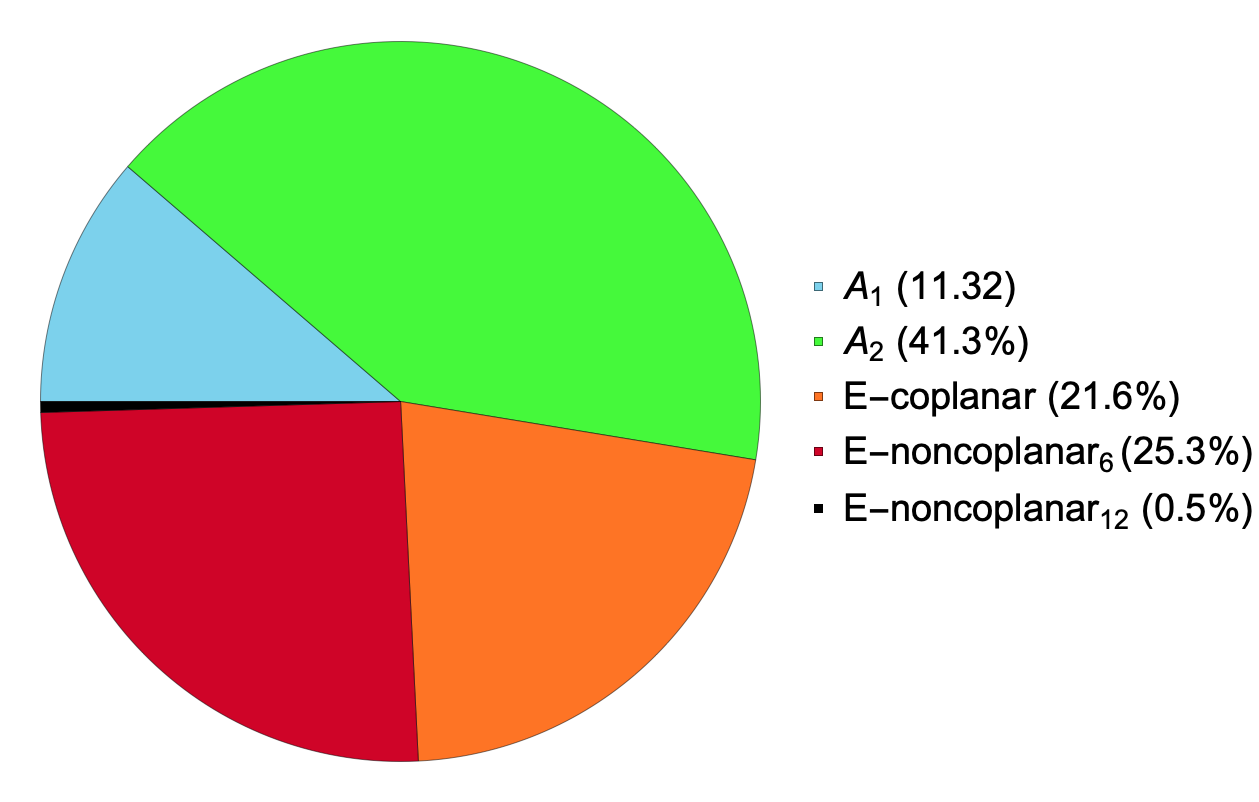}}\\
\subfigure[]{
\includegraphics[width=\columnwidth]{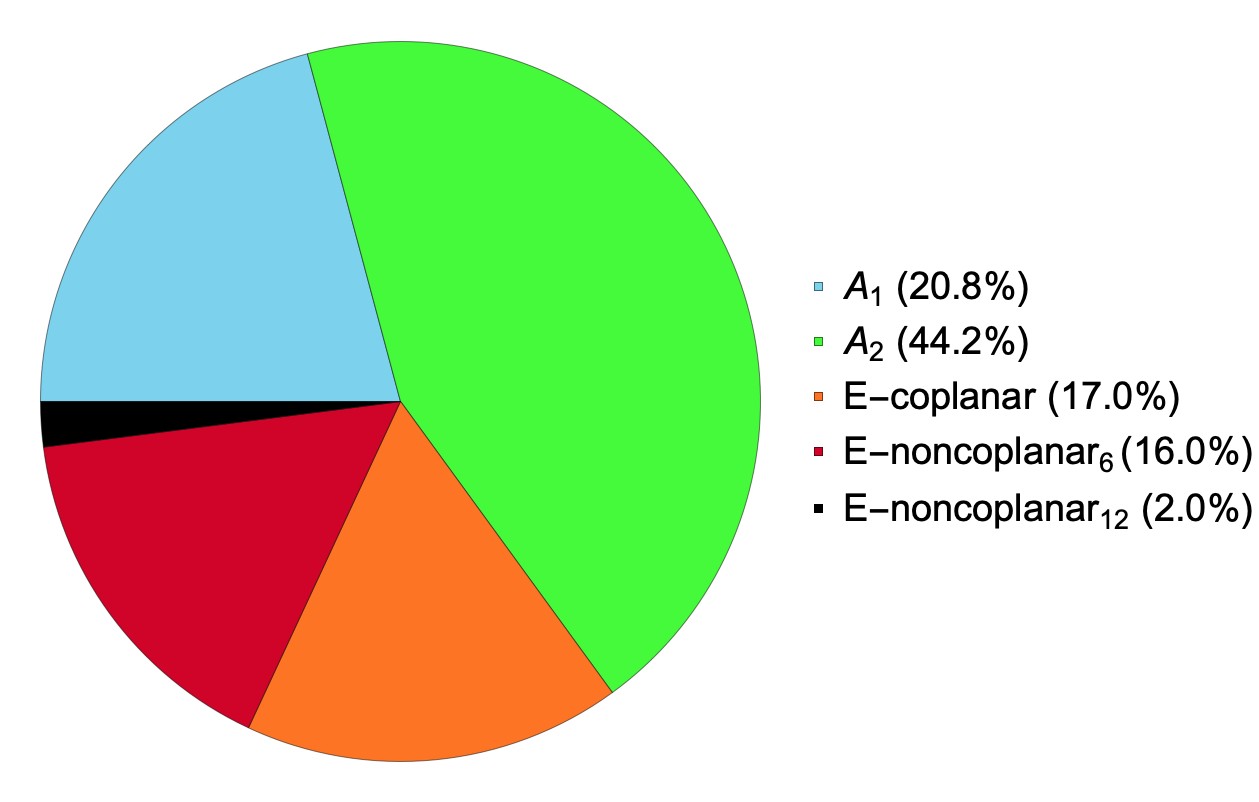}}
\caption{
Relative frequency of different phases within the
full parameter space of the Hamiltonian [Eq. (\ref{eq:H_general})],
with exchange parameters generated randomly from two different distributions.
(a) Exchange parameters are generated randomly according to a uniform distribution on the surface of the
6-dimensional hypersphere defined by Eq. (\ref{eq:hypersphere}).
(b) Diagonal exchange parameters $J_x, J_y, J_z$  are generated according to a uniform distribution on the surface
of a sphere with unit radius, and off-diagonal $D_y, D_z, K$ exchange parameters are generated independently from a 
 uniform distribution on the surface of a sphere with radius$=0.1$ [Eqs. (\ref{eq:pdist1})-(\ref{eq:pdist2})].
This models the effect of the assumption that the scale of off-diagonal couplings is lower. 
The effect on the distribution of phases is minor overall, although  assuming weaker off-diagonal exchange expands
the size of the rare ${\sf E}$-noncoplanar$_{12}$ from $\sim0.5\%$ to $\sim 2\%$
In each case, frequencies are determined by numerically finding the ground state for 100000
random parameter sets generated according to the stated distributions.
}
\label{fig:piechart}
\end{figure}

To investigate the relative frequency of the different phases in the overall parameter space we have calculated the ground state for $100000$ different parameter sets, randomly chosen from a uniform distribution on the surface of the 6-dimensional hypersphere defined by 
\begin{eqnarray}
J_x^2+J_y^2+J_z^2+D_y^2+D_z^2+K^2=1.
\label{eq:hypersphere}
\end{eqnarray}
The pie chart in Fig. \ref{fig:piechart}(a) shows the relative frequency of each of the five phases obtained
from this procedure.
It confirms that ${\sf E}$-noncoplanar$_{12}$ is indeed a rare phase, found as the ground state
for only $\sim 0.5 \%$ of randomly generated parameter sets.
The four other phases are comparatively common.

This leads us to conclude although the ${\sf E}$-noncoplanar$_{12}$ state does not require perfect fine tuning to be realized in a kagome material 
(i.e. it occupies a finite fraction of parameter space), it is unlikely to be realized serendipitously. 
The other four phases should constitute the classical ground states for the vast majority of kagome materials to which the theory in this paper 
can be applied (i.e. those with nearest-neighbour, anisotropic interactions).

The above assumes a probability distribution of parameter sets which is isotropic in the 6-dimensional space
$(J_x, J_y, J_z, D_y, D_z, K)$.
This may not be the case physically, and indeed it is frequently assumed that the off-diagonal components of the exchange tensor $D_y, D_z, K$
should be smaller than the diagonal ones $J_x, J_y, J_z$.
We have investigated the distribution of ground states under this assumption, by generating $100000$ random parameter sets by choosing 
$J_x, J_y, J_z$ from a uniform distribution on the surface of the unit sphere:
\begin{eqnarray}
J_x^2+J_y^2+J_z^2=1
\label{eq:pdist1}
\end{eqnarray}
and indepently choosing 
$D_y, D_z, K$ from a uniform distribution on the surface of a smaller sphere:
\begin{eqnarray}
D_y^2+D_z^2+K^2=0.1.
\label{eq:pdist2}
\end{eqnarray}
The resulting distribution of ground states is shown in  Fig. \ref{fig:piechart}(b).
The relative frequency of different phases is very similar to that with an isotropic distribution of parameters, although the prevalence of the
 ${\sf E}$-noncoplanar$_{12}$ phase increases from $\sim0.5/\%$ to  $\sim2/\%$.

\subsection{Phase diagram in the vicinity of the Antiferromagnetic Heisenberg limit}
\label{subsec:near_heisenberg}

The limit $J_x=J_y=J_z=J>0$, $D_y=K=D_z=0$, gives the well studied nearest neighbor antiferromagnetic
Heisenberg model, which is known to have a highly degenerate ground state \cite{reimers91}.
Generic perturbations away from this limit lift the degeneracy, stabilizing a ground state which is unique up to global
symmetry operations.

\begin{figure*}
\centering
\includegraphics[width=0.06\textwidth]{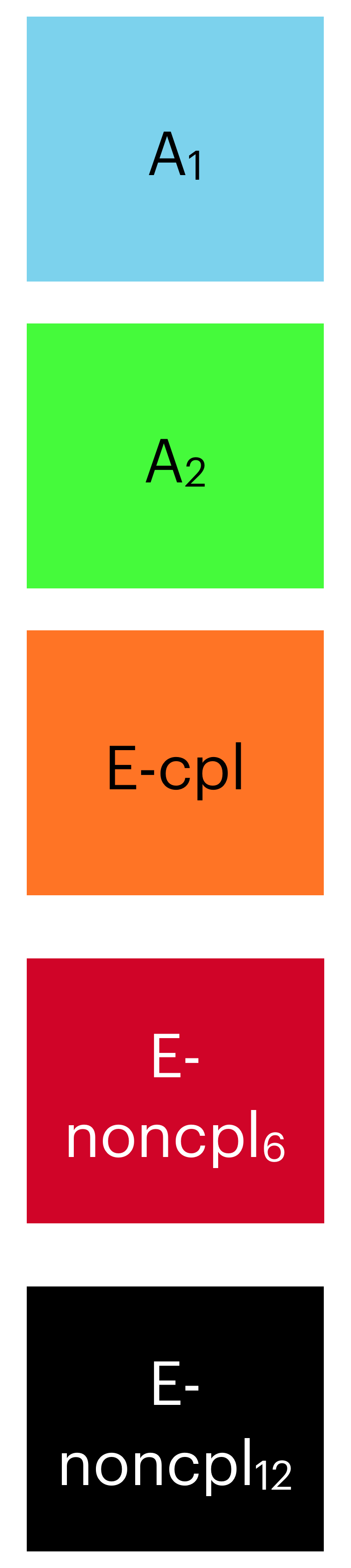}
\subfigure[\ $D_z=-0.25 J$ ]{
\includegraphics[width=0.3\textwidth]{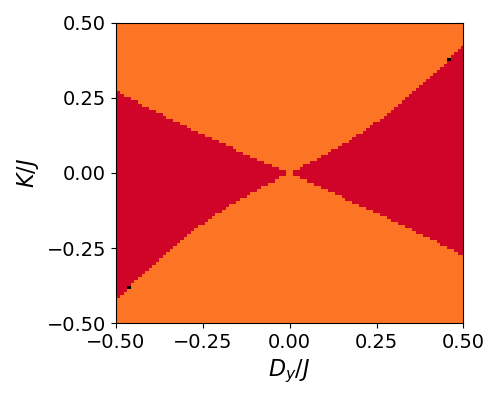}}
\subfigure[\ $D_z=0$  ]{
\includegraphics[width=0.3\textwidth]{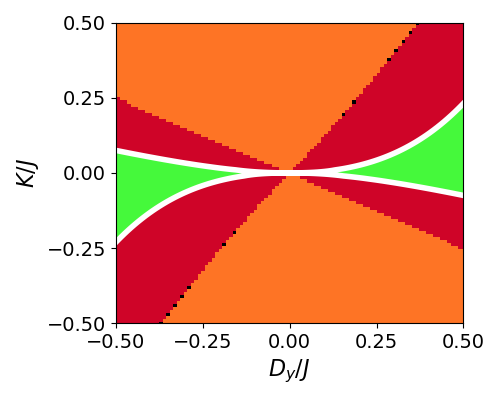}}
\subfigure[\ $D_z=0.25 J$ ]{
\includegraphics[width=0.3\textwidth]{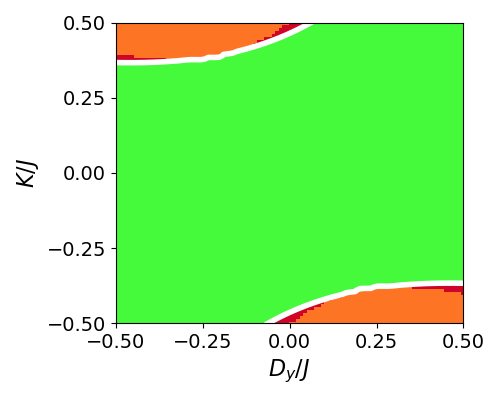}}
\caption{
Ground state phase diagram obtained from perturbing the antiferromagnetic Heisenberg model ($J_x=J_y=J_z=J>0$)
with off-diagonal couplings $D_y, D_z, K$. Phase diagrams are shown as a function of $D_y/J, K/J$ at fixed values of $D_z/J=-0.25$ [(a)],
 $D_z/J=0$ [(b)],  $D_z/J=0.25$ [(c)]
The phase diagram is obtained by comparing numerically optimized energies for the five phases described
in Section \ref{sec:gs}.
The numerical optimization procedure is described in Appendix \ref{app:numerics}.
The $\sf A_1$ phase does not appear on these phase diagrams, as it can only be stabilized as a unique ground state when $J_x<J_y$,
whereas $J_x=J_y$ here.
The white lines show analytic calculations of the boundaries of the $\sf A_2$
phase, using condition (\ref{eq:a2condition}).
}
\label{fig:pd_DyK_nearHeisenberg}
\end{figure*}

Fig. \ref{fig:pd_DyK_nearHeisenberg} shows the effect of perturbing the Heisenberg model with finite off-diagonal
couplings $D_y, D_z, K$. 
$D_z>0$ strongly favours $\sf A_2$ order, while $D_z<0$ favours ordering into  the ${\sf E}$-{coplanar} or ${\sf E}$-{noncoplanar}$_6$ phases depending on which of $D_y$ or $K$ is the more dominant perturbation.
Our results are in agreement with those of Elhajal {\it et al} \cite{Elhajal2002}, who considered the case of perturbing the Heisenberg model with Dzyaloshinskii-Moriya interactions $D_y, D_z$, fixing $K=0$.

When comparing the results here with those of [\onlinecite{Elhajal2002}] one should note that the ground state configurations of the  ${\sf E}-{\rm noncoplanar}_6$  phase become coplanar in the limit of strong positive $J$ and $K=0$. This agrees with the labelling of the same phase as coplanar in [\onlinecite{Elhajal2002}]. Once all symmetry allowed couplings (particularly $K$) are present, this phase becomes non-coplanar, as identified here.

It is notable that the $\sf A_1$ phase does not appear at all in Fig. \ref{fig:pd_DyK_nearHeisenberg}. This can be readily understood from the couplings in Eqs. (\ref{eq:lambdaA1}-\ref{eq:lambdaEcc}).
When $J_x=J_y$, $\lambda_{\sf A_1}=\lambda_{\sf A_2, bb}$.
This then implies that $\omega_{{\sf A_2} 0}\leq\lambda_{\sf A_1}$ [cf. Eqs. (\ref{eq:eval}), (\ref{eq:a1condition})] with the equality only applying when $\lambda_{\sf A_2, ab}=2(\sqrt{3}D_y+K)=0$.

Thus, when $J_x=J_y$ the ${\sf A_2}$ phase will quite generally have a lower energy than the ${\sf A}_1$ phase.
A necessary (but not sufficient) condition for the ${\sf A}_1$ configurations to be the sole ground states
is that $\lambda_{\sf A_1}<\lambda_{\sf A_2 bb} \implies J_x<J_y$.

\begin{figure*}
\centering
\includegraphics[width=0.06\textwidth]{color_key2.pdf}
\subfigure[\ $\delta J_{\perp}=-0.25 J$ ]{
\includegraphics[width=0.3\textwidth]{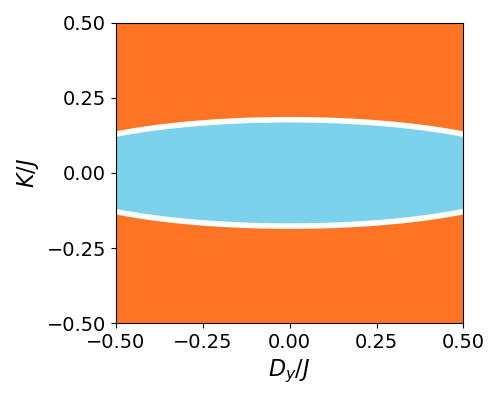}}
\subfigure[\ $\delta J_{\perp}=0.25 J$ ]{
\includegraphics[width=0.3\textwidth]{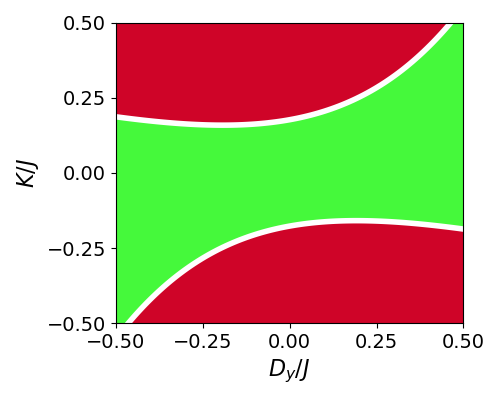}}
\caption{
Ground state phase diagram obtained from perturbing the antiferromagnetic Heisenberg model 
with off-diagonal couplings $D_y, K$, and anisotropy in the transverse exchange $J_x=J+\frac{\delta J_{\perp}}{2}, J_y=J-\frac{\delta J_{\perp}}{2}$.
We set $J_z=J>0$ and $D_z=0$ in both panels.
Phase diagrams are shown as a function of $D_y/J, K/J$ at fixed values of $\delta J_{\perp}/J=-0.25$ [(a)],
$\delta J_{\perp}/J=0.25$ [(b)]
The phase diagram is obtained by comparing numerically optimized energies for the five phases described
in Section \ref{sec:gs}.
The numerical optimization procedure is described in Appendix \ref{app:numerics}.
The white lines show analytic calculations of the boundaries of the $\sf A_1$ and $\sf A_2$
phases, using conditions (\ref{eq:a1condition})-(\ref{eq:a2condition}).
}
\label{fig:pd_DyK_nearHeisenbergdJ}
\end{figure*}

The effect of allowing small anisotropy in the transverse exchanges $J_x, J_y$ is illustrated in Fig. \ref{fig:pd_DyK_nearHeisenbergdJ}.
Here we set
$$D_z=0, \ J_z=J>0, \ J_x=J+\frac{\delta J_{\perp}}{2}, \ J_y=J-\frac{\delta J_{\perp}}{2}
$$
and vary $D_y/J$ and $K/J$.
As implied by the discussion above, $\delta J_{\perp}<0$ favours ${\sf A}_1$ order, becoming unstable to the ${\sf E}$-coplanar phase on increasing $K$.
Conversely, when $\delta J_{\perp}>0$ favours ${\sf A}_2$ order, which gives way to the ${\sf E}$-noncoplanar$_6$ phase for strong $K$.

\section{Relevance to kagome materials}
\label{sec:experiment}

In this section we discuss the application of our results to real
kagome materials.
We divide our discussion into two areas: firstly, rare-earth magnets belonging
to the family R$_3$A$_2$Sb$_3$O$_{14}$ \cite{sanders16, sanders16-jmmc, scheie16,
Dun16a, scheie16, Dun17a, scheie18, ding18, scheie-arXiv19} (sometimes referred to as 
``tripod kagome'' materials \cite{Dun16a, Dun17a}), and secondly,
Cu, Fe and Cr based magnets where exchange anisotropy should be weaker
but nevertheless plays a role in ground state selection.

Aside from the systems mentioned below, we anticipate that ongoing work in synthesizing
frustrated magnets with strong spin-orbit coupling will reveal new kagome systems to which
our results can be applied in the coming years.

\subsection{$\text{R}_3\text{A}_2\text{Sb}_3\text{O}_{14}$ family}
\label{subsec:tripkag}

In the last few years several rare-earth kagome materials with
the general formula $\text{R}_3\text{A}_2\text{Sb}_3\text{O}_{14}$ have
been synthesized. 
This includes materials with A=Mg, Zn and R=Pr, Nd, Sm, Eu, Gd, Tb,
Dy, Ho, Er, Tm, Yb \cite{sanders16, sanders16-jmmc, Dun17a, ding18}.

Where R is a non-Kramers ion (Pr, Eu, Tb, Ho, Tm), the crystal electric
field (CEF) will generally have a non-magnetic singlet ground state, due to
the low symmetry of the rare-earth environment. If the
gap between this singlet and higher CEF states is smaller than or comparable
to the energy scale of interactions, interesting physics may ensue. If the CEF
gap is large, the overall ground state of the system will be a trivial singlet driven
by the onsite physics.
Either way, Eq. (\ref{eq:H_general}) cannot describe such physics without being
augmented by additional terms, so we will not discuss non-Kramers materials
further here.

Where $R$ is a Kramers ion, the CEF will split the 
2J+1 multiplet into a series of doublets. 
At energy and temperature scales below the gap between
the lowest and first excited doublet, the magnetism may be
represented by pseudospin-$1/2$ operators ${\bf S}_i$.
${\bf S}_i$ does not correspond precisely to the magnetic moment,
but relates to it via the $g$-tensor [Eq. \ref{eq:gtensor}].
The important thing for our purposes is that  ${\bf S}_i$
transforms like a magnetic moment with respect to time-reversal
and lattice symmetries, in which case Eqs. (\ref{eq:H_general})-(\ref{eq:J02})
describe the exchange interactions.
Below we briefly discuss the various members of the
$\text{R}_3\text{A}_2\text{Sb}_3\text{O}_{14}$ family, with Kramers ions
$R$, in the 
light of the predictions made in this Article.

The scalar chiral order observed in Nd$_3$Mg$_2$Sb$_3$O$_{14}$ \cite{scheie16, scheie-arXiv19}
corresponds precisely to the ${\sf A_2}$ phase predicted in this work.
The magnetic order of the sister compound Nd$_3$Zn$_2$Sb$_3$O$_{14}$ has not yet been
characterized, but given its essentially similar thermodynamic properties \cite{Dun17a} and crystal field
environment \cite{scheie18} it seems likely to fall in the same phase as Nd$_3$Mg$_2$Sb$_3$O$_{14}$.

 Er$_3$Mg$_2$Sb$_3$O$_{14}$ was reported in Ref. \onlinecite{Dun17a} to avoid long range order
down to very low temperatures. 
It thus appears to be a candidate spin liquid material.
The regions near the phase boundaries of the classical phase diagram presented here are likely
to be particularly fertile ground for the formation of spin liquid states, and this will be an interesting
direction for future research.
 Er$_3$Zn$_2$Sb$_3$O$_{14}$ exhibits strong structural disorder and associated glassy behavior
of the magnetic properties \cite{Dun17a}, which is beyond the scope of our present discussion.

Yb$_3$Mg$_2$Sb$_3$O$_{14}$ exhibits long range order at $T_N\approx0.88$K \cite{Dun17a}.
The form of this magnetic order has yet to be reported in the literature. Based on the expectation that, as a rare
earth magnet with moderate magnetic moment, the theory in this manuscript should be applicable to
Yb$_3$Mg$_2$Sb$_3$O$_{14}$,  we expect that 
the order will be one of the states discussed in this work.
Like Er$_3$Zn$_2$Sb$_3$O$_{14}$, Yb$_3$Zn$_2$Sb$_3$O$_{14}$ has strong structural disorder,
although unlike the Er compound it does not show clear signs of spin freezing \cite{Dun17a}.

Sm$_3$Mg$_2$Sb$_3$O$_{14}$ \cite{sanders16} and Sm$_3$Zn$_2$Sb$_3$O$_{14}$ \cite{sanders16-jmmc}
have both been synthesized but their low temperature magnetism has yet to be characterized in detail.
This may be challenging due to the small magnetic moment of the Sm$^{3+}$ ion, but recent experiments
on the pyrochlores Sm$_2$Ti$_2$O$_7$  and Sm$_2$Sn$_2$O$_7$ indicate that this is possible \cite{pecanha19}.
There is some evidence of hysteresis in the low temperature magnetization curve
for Sm$_3$Zn$_2$Sb$_3$O$_{14}$ \cite{sanders16-jmmc} but not for Sm$_3$Mg$_2$Sb$_3$O$_{14}$ \cite{sanders16},
which may provide some clue as to the low temperature state.

Materials with R=Gd present a somewhat different case, because Hund's rules imply vanishing orbital angular momentum
$L=0$ for the Gd$^{3+}$ ion.
The magnetism on the Gd sites thus comes from a pure $S=7/2$ spin and anisotropies in the interactions should be much weaker.
Some understanding of this case can be gained from considering a model with nearest neighbor Heisenberg exchange and
the nearest-neighbor part of the dipolar interaction: 
\begin{eqnarray}
\mathcal{H}=J \sum_{\langle ij \rangle} {\bf S}_i \cdot {\bf S}_j + \tilde{D}_{\sf nn} \sum_{\langle ij \rangle} \left( {\bf S}_i \cdot {\bf S}_j - 3  {\bf S}_i\cdot\hat{\bf r}_{ij} {\bf S}_j\cdot\hat{\bf r}_{ij} \right) \nonumber \\
\label{eq:hdip}
\end{eqnarray}
In terms of the symmetry-allowed interaction matrices [Eqs. (\ref{eq:J01})-(\ref{eq:J02})]
this Hamiltonian corresponds to setting
\begin{eqnarray}
&&J_x=J-2 \tilde{D}_{\sf nn}, \ 
J_y=J_z=J+\tilde{D}_{\sf nn}, \nonumber \\
&&D_y=D_z=K=0.
\label{eq:dip-coeffs}
\end{eqnarray}
Inserting Eq. (\ref{eq:dip-coeffs}) into Eqs. (\ref{eq:lambdaA1})-(\ref{eq:lambdaEbc}) leads us 
to the conclusion that for $J, \tilde{D}_{\sf nn}>0$, the $A_1$ configuration is favored out
of the forms of order considered in this Article.
This agrees
with the conclusions of Maksymenko {\it et al} \cite{maksymenko15}, who studied
the phase diagram incorporating isotropic nearest neighbor exchange $J$ with the full long ranged
dipolar interaction $D$, and found the $A_1$ configuration as the ground state for
weak to moderate $D$ and antiferromagnetic $J$. 
It also agrees with previous predictions about the ground state of Gd$_3$Mg$_2$Sb$_3$O$_{14}$ 
\cite{Dun16a}, and with the observed antiferromagnetic transition at $T_N\approx1.7 K$ \cite{Dun16a, wellm20}, 
although differences between the field cooled and zero-field cooled susceptibility \cite{wellm20} remain to be understood.

For R=Dy the ionic magnetic moment is very large and the long
range component of the dipolar interaction cannot be ignored.
Dy$_3$Mg$_2$Sb$_3$O$_{14}$ exhibits an unusual ``fragmented'' \cite{brooks14a} phase
where there is an ordering of emergent ``charge''
degrees of freedom while spins remain partially disordered \cite{Paddison16b}.
The long-range dipole-dipole interaction plays a crucial role
in this phenomenon \cite{moller09, chern11} and thus it is beyond the scope of the
theory presented in this Article.

\subsection{Nearly isotropic systems}
\label{subsec:nearly_isotropic}

While the most obvious application of the results in this Article is found in systems where
exchange anisotropy is strong, our results can also be applied to understand cases where
isotropic Heisenberg exchange is weakly perturbed by short ranged anisotropic interactions.
%

This is the case in the Fe- and Cr- jarosites AM$_3$(OH)$_6$(SO$_4$)$_2$ where M= \{Fe, Cr\} and  
A=\{K, Rb, NH$_4$, Na\}\cite{grohol03, nishiyama03, morimoto03, Matan06, yildirim06}. 
These are found to order in the $A_2$ phase - the most prevalent of our phase diagram.
This is generally understood to be a consequence of antiferromagnetic Heisenberg exchange
perturbed by a weak $D_y$.
This interpretation fully agrees with the results presented here: it can readily be checked that inserting
\begin{eqnarray}
&&J_x=J_y=J_z=J>0 \nonumber \\
&&D_z=K=0, \ \ |D_y|<<J
\end{eqnarray}
into Eqs. (\ref{eq:lambdaA1})-(\ref{eq:lambdaEbc}) gives an outcome obeying condition (\ref{eq:a2condition})
and hence a ground state in the ${\sf A_2}$ phase [cf. Fig. \ref{fig:pd_DyK_nearHeisenberg}].
What this work adds to the discussion is a simple and systematic approach to finding the preferred
ground state for general kinds of anisotropic nearest neighbor perturbation.

An example where weak anisotropic perturbations away from a Heisenberg model lead to something
other than ${\sf A_2}$ order is given by Cd-kapellasite \cite{okuma17}.
The weak ferromagnetic moment confined within the kagome planes in that material is 
only consistent with the ${\sf E}$-coplanar phase, out of the phases in this Article.


\section{Summary and Discussion}
\label{sec:conclusions}

In this Article we have developed a theory of the magnetic orders induced by nearest-neighbor exchange
anisotropy in kagome magnets.
Our theory reveals that five distinct magnetic orders can be expected from such interactions, all retaining
the translational symmetry of the lattice, but being distinguished from one another by their transformations
under time-reversal and point group symmetries.
The five phases are: ${\sf A}_1$ [Fig. \ref{fig:A1}], 
${\sf A}_2$ [Fig. \ref{fig:A2}], 
${\sf E}$-{coplanar}  [Fig. \ref{fig:Ecoplanar}], 
${\sf E}$-{noncoplanar}$_6$  [Fig. \ref{fig:Enoncoplanar6}], 
${\sf E}$-{noncoplanar}$_{12}$  [Fig. \ref{fig:Enoncoplanar12}].
They are labelled according to the irreducible representation of the point group $C_{3v}$ with which the
primary order parameter transforms, their coplanar or noncoplanar nature and their degeneracy.
Eqs. (\ref{eq:a1condition})-(\ref{eq:a2condition}) give exact conditions for the ${\sf A_1}$ and ${\sf A_2}$
configurations to be classical ground states.

We have used numerical calculations to determine the full zero temperature phase diagram of the most
general anisotropic nearest-neighbor exchange model, showing the extent of these five phases 
[Figs. \ref{fig:pd_j+_k-050}-\ref{fig:pd_j-_k+050}].
One of the five phases (${\sf E}$-{noncoplanar}$_{12}$) is found to be exceedingly rare in the parameter space [Fig. \ref{fig:piechart}].

We have discussed how this theory relates to various real kagome materials [Section \ref{sec:experiment}], with both strong and weak exchange anisotropy.

The dominance of noncollinear (${\sf A}_1$, ${\sf E}$-coplanar) and noncoplanar ($A_2$, ${\sf E}$-noncoplanar$_{6, 12}$)
states on the phase diagram suggests a high possibility of spin excitations with topological band structures in many kagome materials \cite{owerre17,
seshadri18, mook19}.
It is likely that the five phases identified here from analysis of broken symmetries can be subdivided further by the topology of
the excitation bands.
Relatedly, the possibility of coupling to itinerant electrons is
an interesting area for future research with a view to investigating topological transport
phenomena.

The approach used in this work relies on the ability to decompose the Hamiltonian into a sum over blocks, such that the ground state is obtained by finding the ground state on each block and tiling it over the lattice. This would seem to limit the usefulness of the approach for
systems with further neighbor interactions, since such a decomposition may either not be possible or may require such large blocks that the decomposition is no longer a useful simplification.
Applying the method from this work to quantum systems will also not be possible in general - even for nearest neighbor interactions - because the Hamiltonians on neighboring blocks will usually not commute.
There are, however, some specific, fine-tuned, cases where the exact ground state of a quantum system can be built up by such a block-by-block
approach \cite{changlani18, palle21}.

While we have restricted ourselves here to phases which are stable 
over finite regions of the classical phase diagram, a study of the phase boundaries
may also be interesting. 
As has been studied elsewhere \cite{essafi17, yan17} phase
boundaries between competing classical phases can host non-trivial enlarged manifolds
of zero-energy states, which in some cases are associated with new forms of spin liquid \cite{benton16-pinchline}.
In general, the greater the degree of degeneracy around the phase boundary, the more
more favorable the situation becomes towards the formation of spin liquids.
Different phase boundaries will have different amounts of additional degeneracy and so some
will be more favorable for spin liquid formation than others. 
Boundaries where 3 (rather than just 2) phases meet may host particularly interesting physics as seen in (e.g.) [\onlinecite{benton16-pinchline}].
An analysis of each possible phase boundary would be an interesting undertaking, which we leave open 
for future work.

\section*{Acknowledgements}
The author thanks  Zhiling Dun, Karim Essafi, Ludovic Jaubert and Han Yan for 
helpful discussions and collaborations on related work.
Karim Essafi, Ludovic Jaubert and Johannes Richter are also thanked for feedback on the draft 
manuscript.
The author  acknowledges the hospitality of LOMA at the University of Bordeaux,
where part of this work was carried out.

\appendix

\section{Numerical optimization of energies}
\label{app:numerics}

Here we describe the numerical optimization used to 
obtain the phase diagrams in Figs. \ref{fig:pd_j+_k-050}-\ref{fig:pd_j-_k+050}
and the estimates of the relative frequency of phases in Fig. \ref{fig:piechart}.

For a given parameter set, the energy is optimized separately for each of the five
phases described in Section \ref{sec:gs} and then the optimized energies are compared to
determine which is the lowest.

Due to the argument in Section \ref{subsec:q=0}, we need only optimize the configuration
on a single triangle, since we know that a ground state on the full lattice can be obtained
by tiling the ground state of a single triangle everywhere.

The optimization for each phase is done by either random search or simulated annealing
combined with iterative minimization \cite{sim18}, apart from 
the $A_1$ phase where the spin configuration
is fixed [Eq. \ref{eq:A1config}] and thus the corresponding energy can directly be calculated without any optimization being necessary:
\begin{eqnarray}
E_{\sf A_1}=\frac{3}{4}\left( -2 \sqrt{3} D_z + J_x -3 J_y \right).
\label{eq:A1}
\end{eqnarray}

For the other four phases (${\sf A}_2$, ${\sf E}$-coplanar,  ${\sf E}$-noncoplanar$_6$, 
 ${\sf E}$-noncoplanar$_{12}$), the optimization procedure is as described below.

\subsection{Optimizing ${\sf A_2}$ configuration}
\label{app:a2optimize}

The form for the $\sf A_2$ configurations is given in Eq. (\ref{eq:A2config}).
This can be written as 
\begin{eqnarray}
&&{\bf S}_0=\left( -\frac{\sqrt{3}}{2} s_a, s_a/2, s_b \right) \\
&&{\bf S}_1=\left( \frac{\sqrt{3}}{2} s_a, s_a/2, s_b \right) \\
&&{\bf S}_2=\left(0, -s_a, s_b \right)
\end{eqnarray}
with $(s_a, s_b)$ on the unit circle
\begin{eqnarray}
s_a^2+s_b^2=1.
\label{eq:unit_circle_s}
\end{eqnarray}

Initially, we calculate the energy for $10^5$ randomly generated values
of $(s_a, s_b)$ on the unit circle.
The lowest energy configuration obtained from this random search is then
used as input for the iterative minimization step.

In the iterative minimization step $(s_a, s_b)$ are updated as
\begin{eqnarray}
s_a \to \frac{s_a-c \frac{\partial E}{\partial s_a}}{|(s_a-c \frac{\partial E}{\partial s_a}, 
s_b-c \frac{\partial E}{\partial s_b})|}  \nonumber \\
s_b \to \frac{s_b-c \frac{\partial E}{\partial s_b}}{|(s_a-c \frac{\partial E}{\partial s_a}, 
s_b-c \frac{\partial E}{\partial s_b})|}.
\label{eq:a2update}
\end{eqnarray}
For sufficiently small, positive, $c$ this update is guaranteed to reduce the energy, unless the 
system is already in a locally optimal configuration before the update.

The parameter $c$ is initially set to $0.1$. If the update (\ref{eq:a2update}) does not
reduce the energy then $c$ is reduced by a factor of $2$ and the update
is attempted again.
This procedure is repeated until the configuration converges.

\subsection{Optimizing ${\sf E}$-coplanar configuration}
\label{app:ecploptimize}

The form for an $\sf E$-coplanar configuration is given in Eq. (\ref{eq:cpl-config}).
This can be rewritten as
\begin{eqnarray}
&&{\bf S}_0=\left( \sigma_x, \sigma_y, \sigma_z \right) \\
&&{\bf S}_1=\left( \sigma_x, -\sigma_y, -\sigma_z \right) \\
&&{\bf S}_2=\left(1,0,0 \right)
\end{eqnarray}
with  $( \sigma_x, \sigma_y, \sigma_z)$ on the unit sphere
\begin{eqnarray}
\sigma_x^2+\sigma_y^2+\sigma_z^2=1.
\label{eq:unit_circle_sigma}
\end{eqnarray}

Initially, we calculate the energy for $10^5$ randomly generated values
of $(\sigma_x, \sigma_y, \sigma_z)$ on the unit sphere.
The lowest energy configuration obtained from this random search is then
used as input for the iterative minimization step.

In the iterative minimization step $(\sigma_x, \sigma_y, \sigma_z)$ are updated as
\begin{eqnarray}
\sigma_{\alpha} \to \frac{\sigma_{\alpha}-c \frac{\partial E}{\partial \sigma_{\alpha}}}{|(\sigma_x-c \frac{\partial E}{\partial \sigma_x}, 
\sigma_y-c \frac{\partial E}{\partial \sigma_y},
\sigma_z-c \frac{\partial E}{\partial \sigma_z})|}
\label{eq:cplupdate}
\end{eqnarray}

The parameter $c$ is initially set to $0.1$. If the update (\ref{eq:cplupdate}) does not
reduce the energy then $c$ is reduced by a factor of $2$ and the update
is attempted again.
This procedure is repeated until the configuration converges.

The set of configurations covered by the ${\sf E}$-coplanar ansatz (\ref{eq:cpl-config})
includes the ${A_1}$ configurations (when $\phi=\frac{4 \pi}{3}, \theta=\frac{\pi}{2}$).
Because of this, if the 
${\sf E}$-coplanar optimization is found to give the lowest energy of the five possibilities 
we must check that the obtained configuration has a nonzero value of at least one of the order parameters  ${\bf m}_{{\sf E} \alpha}$.
In practice we check that
\begin{eqnarray}
|{\bf m}_{{\sf E} a}|^2+|{\bf m}_{{\sf E} b}|^2+|{\bf m}_{{\sf E} c}|^2>10^{-5}.
\label{eq:cpl-check}
\end{eqnarray}
If the 
${\sf E}$-coplanar optimization obtains the lowest energy but the inequality (\ref{eq:cpl-check})
is not fulfilled, the ground state is assigned to the ${\sf A}_1$ phase.

\subsection{Optimizing ${\sf E}$-noncoplanar$_6$ configuration}
\label{app:enoncpl6optimize}

The form for an $\sf E$-noncoplanar$_6$ configuration is given in Eq. (\ref{eq:cpl-config}).
This can be rewritten as
\begin{eqnarray}
&&{\bf S}_0=\left( \tau_x, \tau_y, \tau_z \right) \\
&&{\bf S}_1=\left( -\tau_x, \tau_y, \tau_z \right) \\
&&{\bf S}_2=\left(0, t_a, t_b \right)
\end{eqnarray}
with  $( \tau_x, \tau_y, \tau_z)$ on the unit sphere and $(t_a, t_b)$ on the unit circle
\begin{eqnarray}
&&\tau_x^2+\tau_y^2+\tau_z^2=1
\label{eq:unit_circle_tau}\\
&&t_a^2+t_b^2=1.
\label{eq:unit_circle_t}
\end{eqnarray}

Initially, we calculate the energy for $10^5$ randomly generated values
of $(\tau_x, \tau_y, \tau_z)$ and $(t_a, t_b)$ obeying 
Eqs. (\ref{eq:unit_circle_tau})-(\ref{eq:unit_circle_t}).
The lowest energy configuration obtained from this random search is then
used as input for the iterative minimization step.

In the iterative minimization step, we update the parameters according to the following:
\begin{eqnarray}
&&\tau_{\alpha} \to \frac{\tau_{\alpha}-c \frac{\partial E}{\partial \tau_{\alpha}}}{|(\tau_x-c \frac{\partial E}{\partial \tau_x}, 
\tau_y-c \frac{\partial E}{\partial \tau_y},
\tau_z-c \frac{\partial E}{\partial \tau_z})|}  \nonumber \\
&&t_{\alpha} \to \frac{t_{\alpha}-c \frac{\partial E}{\partial t_{\alpha}}}{|(t_a-c \frac{\partial E}{\partial t_a}, 
t_b-c \frac{\partial E}{\partial t_b})|} 
\label{eq:noncplupdate}
\end{eqnarray}
The parameter $c$ is initially set to $0.1$. If the update (\ref{eq:noncplupdate}) does not
reduce the energy then $c$ is reduced by a factor of $2$ and the update
is attempted again.
This procedure is repeated until the configuration converges.

The set of configurations covered by the ${\sf E}$-noncoplanar$_6$ ansatz (\ref{eq:noncpl-config})
includes the ${A_2}$ configurations (when $\mu=-(\kappa-\frac{\pi}{2}), \nu=-\frac{\pi}{6}$).
Because of this, if the 
${\sf E}$-noncoplanar$_6$ optimization is found to give the lowest energy of the five possibilities we must check 
that the obtained configuration has a nonzero value of at least one of the order parameters  ${\bf m}_{{\sf E} \alpha}$.
Numerically, we check the condition (\ref{eq:cpl-check}).
If the 
${\sf E}$-noncoplanar$_6$ optimization obtains the lowest energy but the inequality (\ref{eq:cpl-check})
is not fulfilled, the ground state is assigned to the ${\sf A}_2$ phase.

\subsection{Optimizing ${\sf E}$-noncoplanar$_{12}$ configuration}
\label{app:enoncpl12optimize}

Because the ${\sf E}$-noncoplanar$_{12}$ state allows for any configuration of three spins on a single
triangle, the configuration space of states is larger and we use simulated annealing rather than a purely
random search for the initial optimization, before the iterative minimization step.

In the simulated annealing the three spins on a triangle are initialized in a random configuration.
Updates are attempted one spin at a time, being certainly accepted if they reduce the energy and accepted
with probability $\exp(-\delta E/T)$ if they increase the energy by an amount $\delta E$.
Initially, the ``temperature'', $T=0.2$ in units where $|J_z|=1$ (for Figs.  \ref{fig:pd_j+_k-050}- \ref{fig:pd_j-_k+050}] )
or where $J_x^2+J_y^2+J_z^2+D_y^2+D_z^2+K^2=1$ (for Fig. \ref{fig:piechart}(a)) 
or where $J_x^2+J_y^2+J_z^2=1$ (for Fig. \ref{fig:piechart}(b)).
The triangle is swept $10^5$ times at a given temperature, and the temperature is then reduced by a factor
of 0.9.
This procedure is repeated 200 times.
There are than $10^5$ sweeps of the triangle with $T=0$, i.e. only accepting energy reducing updates.

The whole annealing procedure is performed from the start 3 times for each parameter set with the
final output being the lowest energy configuration obtained over all three sweeps.

To converge the configuration further, there is then an iterative minimisation step where each
spin component is updated as:
\begin{eqnarray}
S_i^{\alpha}\to \frac{S_i^{\alpha}-c \frac{\partial E}{\partial S_i^{\alpha}}}{|(S_i^{x}-c \frac{\partial E}{\partial S_i^{x}},
S_i^{y}-c \frac{\partial E}{\partial S_i^{y}},
S_i^{z}-c \frac{\partial E}{\partial S_i^{z}}
)|}
\label{eq:itminupdate}
\end{eqnarray}
The parameter $c$ is initially set to $0.1$. If the update (\ref{eq:itminupdate}) does not
reduce the energy then $c$ is reduced by a factor of $2$ and the update
is attempted again.
This procedure is repeated until the configuration converges.

If the energy produced from this procedure is lower than the energy produced from optimizing within
the ${\sf A}_1$, ${\sf A}_2$, $E$-coplanar or $E$-noncoplanar$_6$ phases, then the ground state
may be within the $E$-noncoplanar$_{12}$ phase.
Because the configuration on the triangle is completely general, to confirm that the configuration has not converged to one
of the other phases we check that the inequality 
(\ref{eq:cpl-check}) is satisfied, and also check that:
\begin{eqnarray}
&&m_{\sf A_1}^2>10^{-5} 
\label{eq:12check1}
\\
&&m_{\sf A_2a}^2+m_{\sf A_2b}^2>10^{-5}.
\label{eq:12check2}
\end{eqnarray}
If inequalities (\ref{eq:cpl-check}), (\ref{eq:12check1}), (\ref{eq:12check2}) are not satisfied, the ground state is
assigned to one of the other phases depending on the values of the various ${\bf m}_{\gamma}$ [Table~\ref{table:ansatz}].

\bibliography{library.bib}

\end{document}


\title{Supplemental Material: Ordered ground states of kagome magnets with generic exchange anisotropy}

\author{Owen Benton}
\affiliation{Max Planck Institute for the Physics of Complex Systems, N{\"o}thnitzer Str. 38, Dresden 01187, Germany}

\maketitle

In this Supplemental Material we present phase diagrams for a series of additional values of $D_y/|J_z|, D_z/|J_z|, K/|J_z|$ from 
$\{-0.75, -0.25, 0.25, 0.75  \}$, with both signs of $J_z$.

\begin{figure*}
\centering
\includegraphics[width=0.5\textwidth]{color_key.pdf}\\
\includegraphics[width=0.8\textwidth]{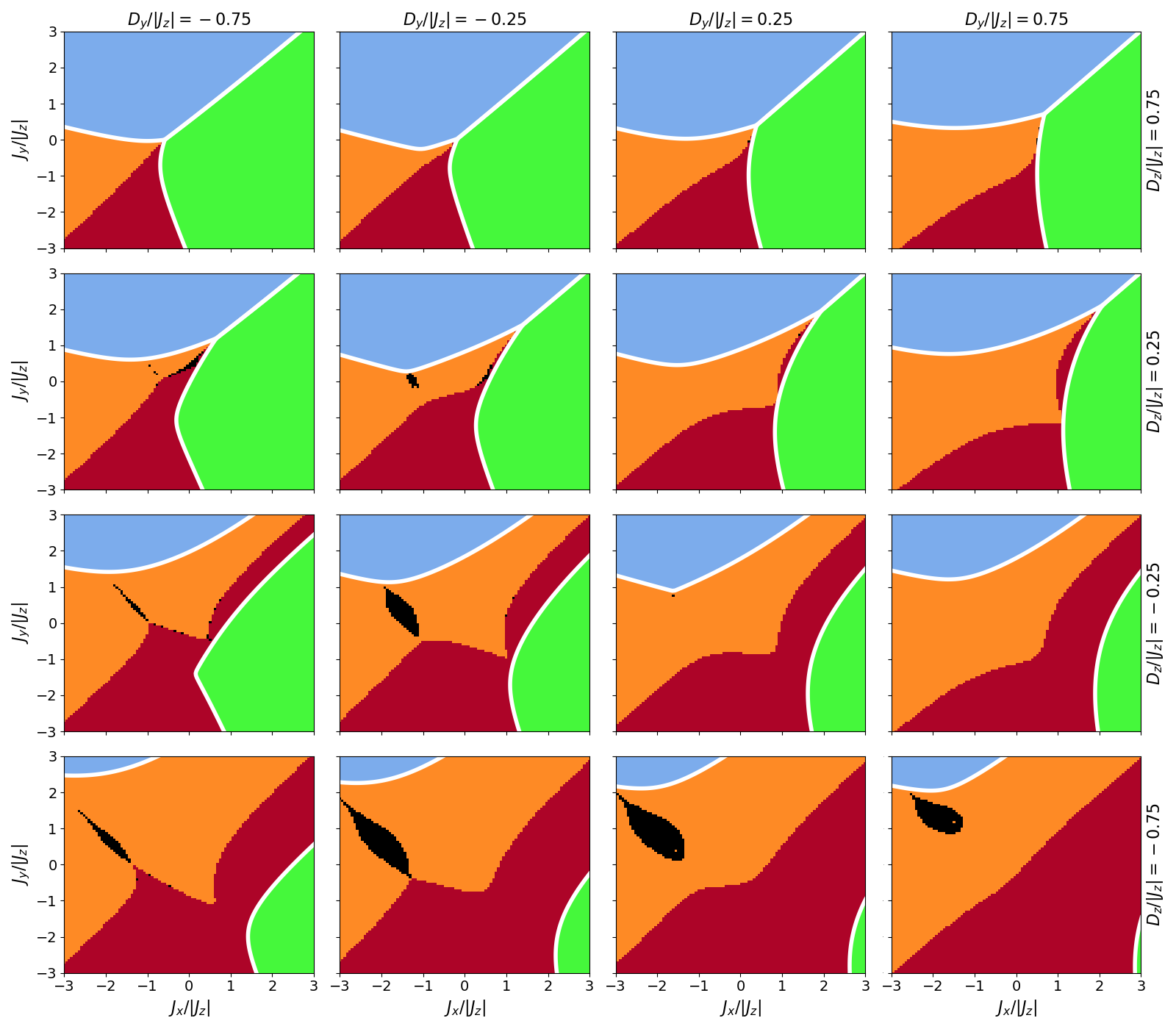}
\caption{$T=0$ phase diagram with $J_z>0$ and $K=-0.75 |J_z|$.
Each panel shows a slice of the phase diagram as a function of $J_x$ and $J_y$ for different, 
fixed, values of the DM directions $D_y$ and $D_z$, with $D_y$ increasing from left to right and
$D_z$ from bottom to top.
The phase diagram is obtained by comparing numerically optimized energies for the five phases described
in the main text.
The white lines show analytic calculations of the boundaries of the $\sf A_1$ and $\sf A_2$
phases.
}
\label{fig:pd_j+_k-075}
\end{figure*}

\begin{figure*}
\centering
\includegraphics[width=0.5\textwidth]{color_key.pdf}\\
\includegraphics[width=0.8\textwidth]{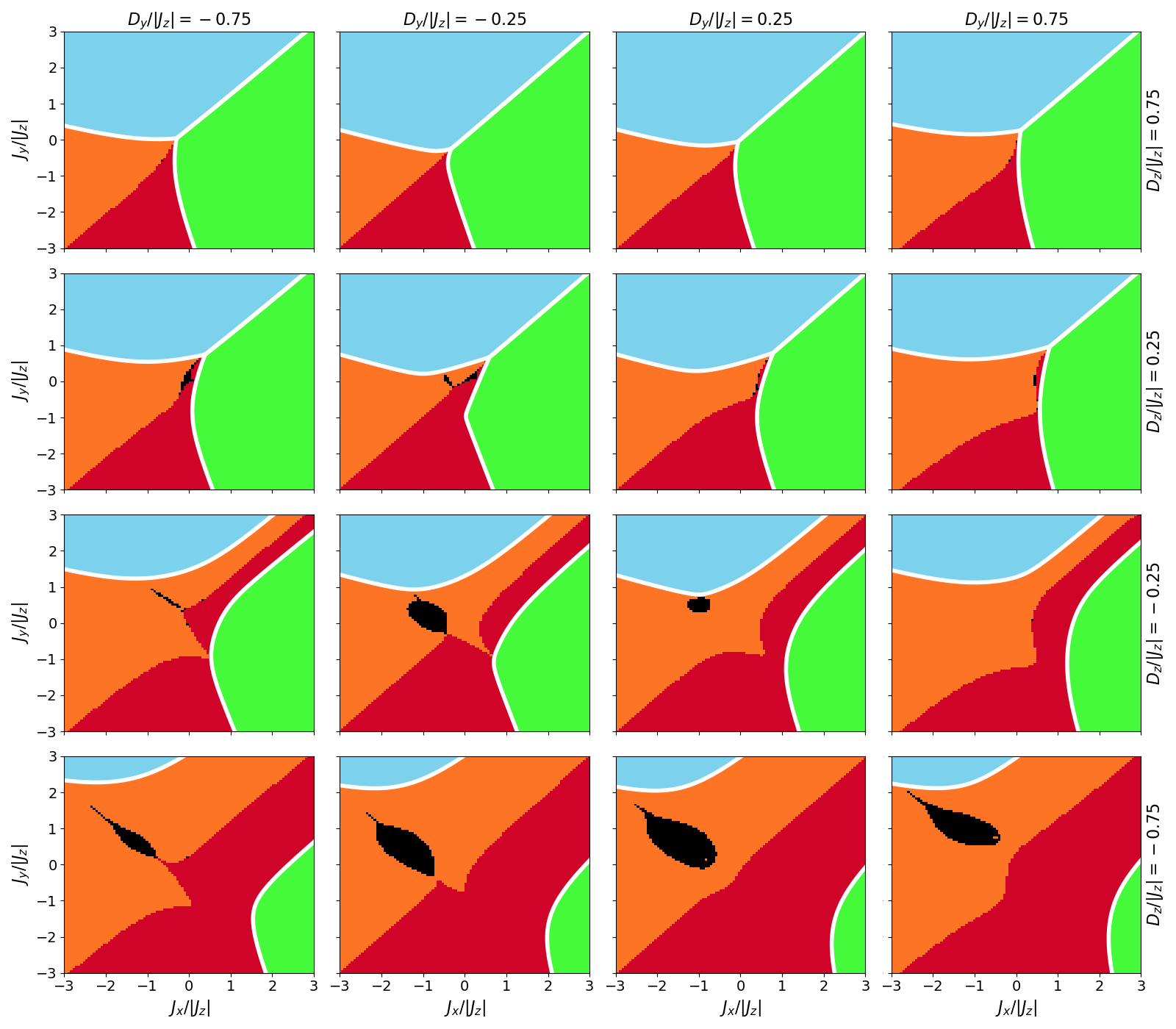}
\caption{$T=0$ phase diagram with $J_z>0$ and $K=-0.25 |J_z|$.
Each panel shows a slice of the phase diagram as a function of $J_x$ and $J_y$ for different, 
fixed, values of the DM directions $D_y$ and $D_z$, with $D_y$ increasing from left to right and
$D_z$ from bottom to top.
The phase diagram is obtained by comparing numerically optimized energies for the five phases described
in the main text.
The white lines show analytic calculations of the boundaries of the $\sf A_1$ and $\sf A_2$
phases.
}
\label{fig:pd_j+_k-025}
\end{figure*}

\begin{figure*}
\centering
\includegraphics[width=0.5\textwidth]{color_key.pdf}\\
\includegraphics[width=0.8\textwidth]{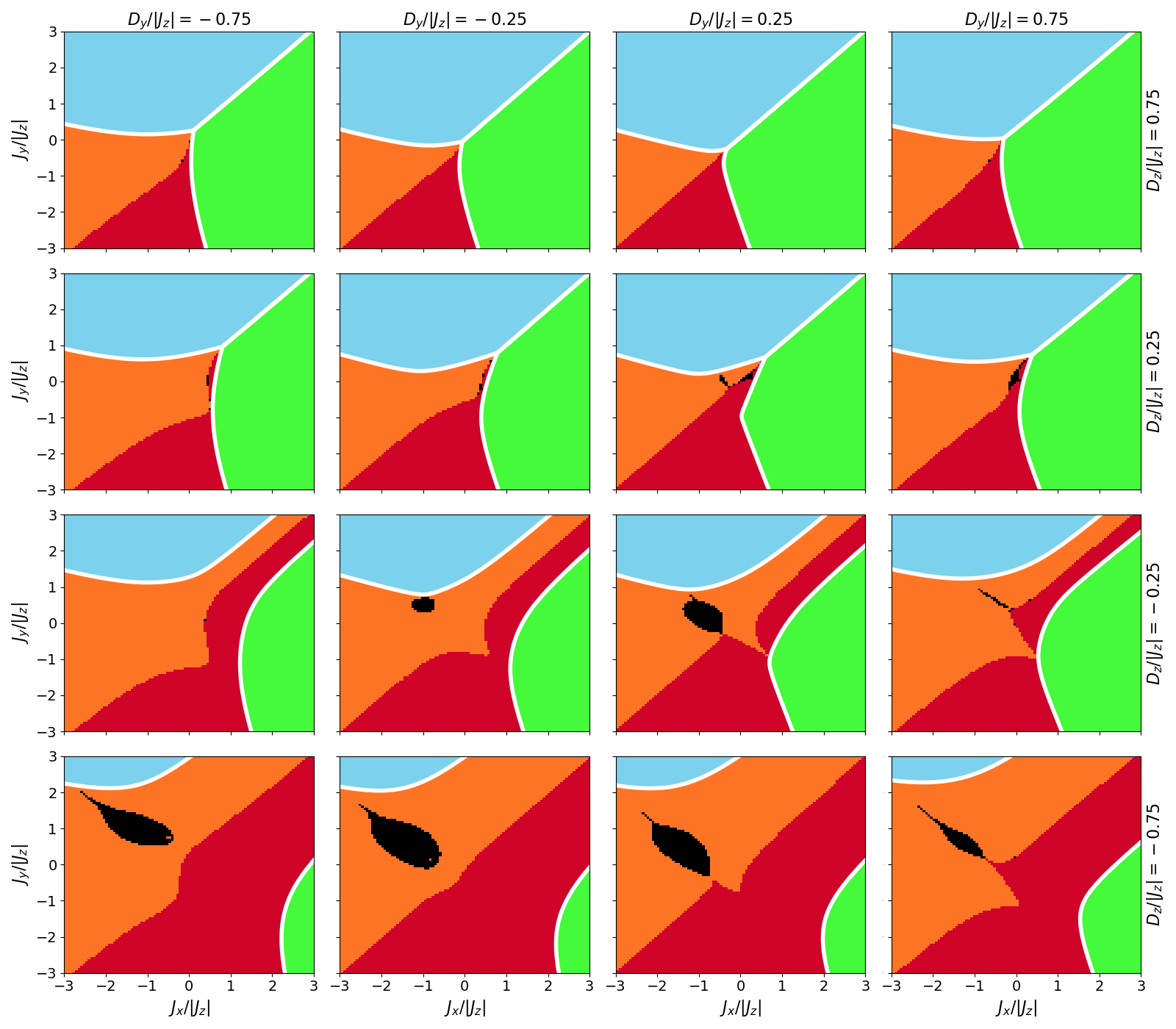}
\caption{$T=0$ phase diagram with $J_z>0$ and $K=0.25 |J_z|$.
Each panel shows a slice of the phase diagram as a function of $J_x$ and $J_y$ for different, 
fixed, values of the DM directions $D_y$ and $D_z$, with $D_y$ increasing from left to right and
$D_z$ from bottom to top.
The phase diagram is obtained by comparing numerically optimized energies for the five phases described
in the main text.
The white lines show analytic calculations of the boundaries of the $\sf A_1$ and $\sf A_2$
phases.
}
\label{fig:pd_j+_k+025}
\end{figure*}

\begin{figure*}
\centering
\includegraphics[width=0.5\textwidth]{color_key.pdf}\\
\includegraphics[width=0.8\textwidth]{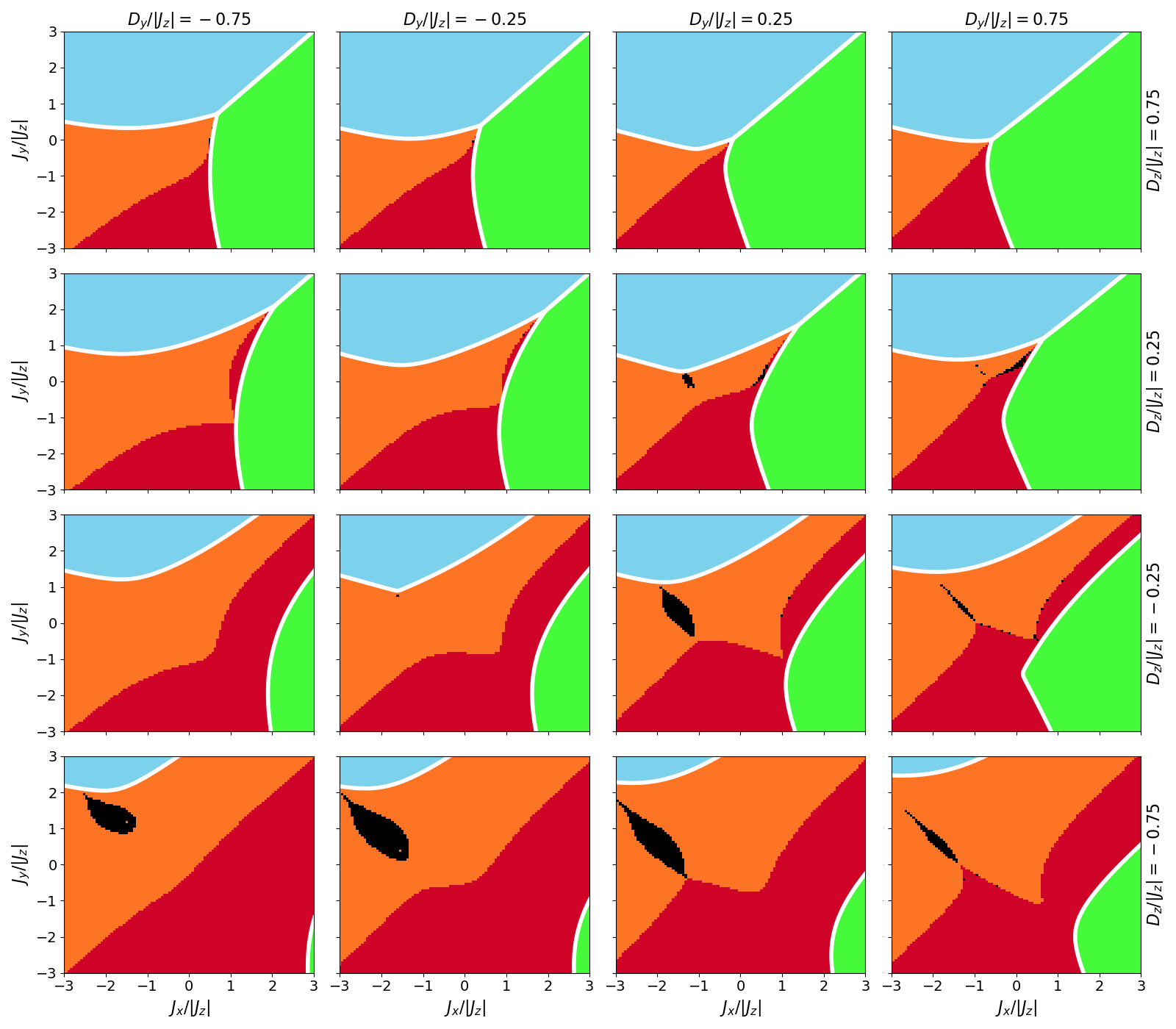}
\caption{$T=0$ phase diagram with $J_z>0$ and $K=0.75 |J_z|$.
Each panel shows a slice of the phase diagram as a function of $J_x$ and $J_y$ for different, 
fixed, values of the DM directions $D_y$ and $D_z$, with $D_y$ increasing from left to right and
$D_z$ from bottom to top.
The phase diagram is obtained by comparing numerically optimized energies for the five phases described
in the main text.
The white lines show analytic calculations of the boundaries of the $\sf A_1$ and $\sf A_2$
phases.
}
\label{fig:pd_j+_k+075}
\end{figure*}

\begin{figure*}
\centering
\includegraphics[width=0.5\textwidth]{color_key.pdf}\\
\includegraphics[width=0.8\textwidth]{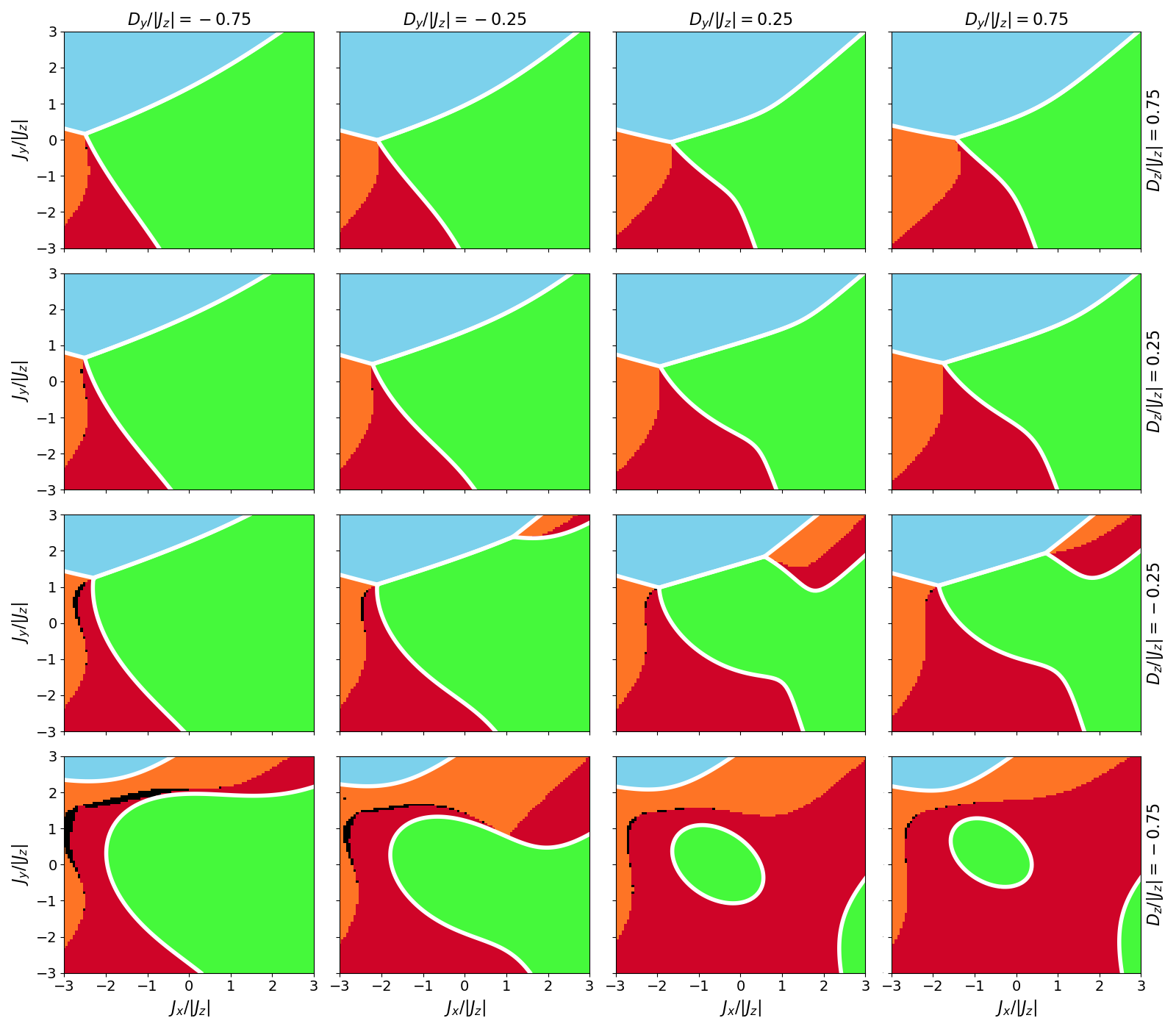}
\caption{$T=0$ phase diagram with $J_z<0$ and $K=-0.75 |J_z|$.
Each panel shows a slice of the phase diagram as a function of $J_x$ and $J_y$ for different, 
fixed, values of the DM directions $D_y$ and $D_z$, with $D_y$ increasing from left to right and
$D_z$ from bottom to top.
The phase diagram is obtained by comparing numerically optimized energies for the five phases described
in the main text.
The white lines show analytic calculations of the boundaries of the $\sf A_1$ and $\sf A_2$
phases.
}
\label{fig:pd_j-_k-075}
\end{figure*}

\begin{figure*}
\centering
\includegraphics[width=0.5\textwidth]{color_key.pdf}\\
\includegraphics[width=0.8\textwidth]{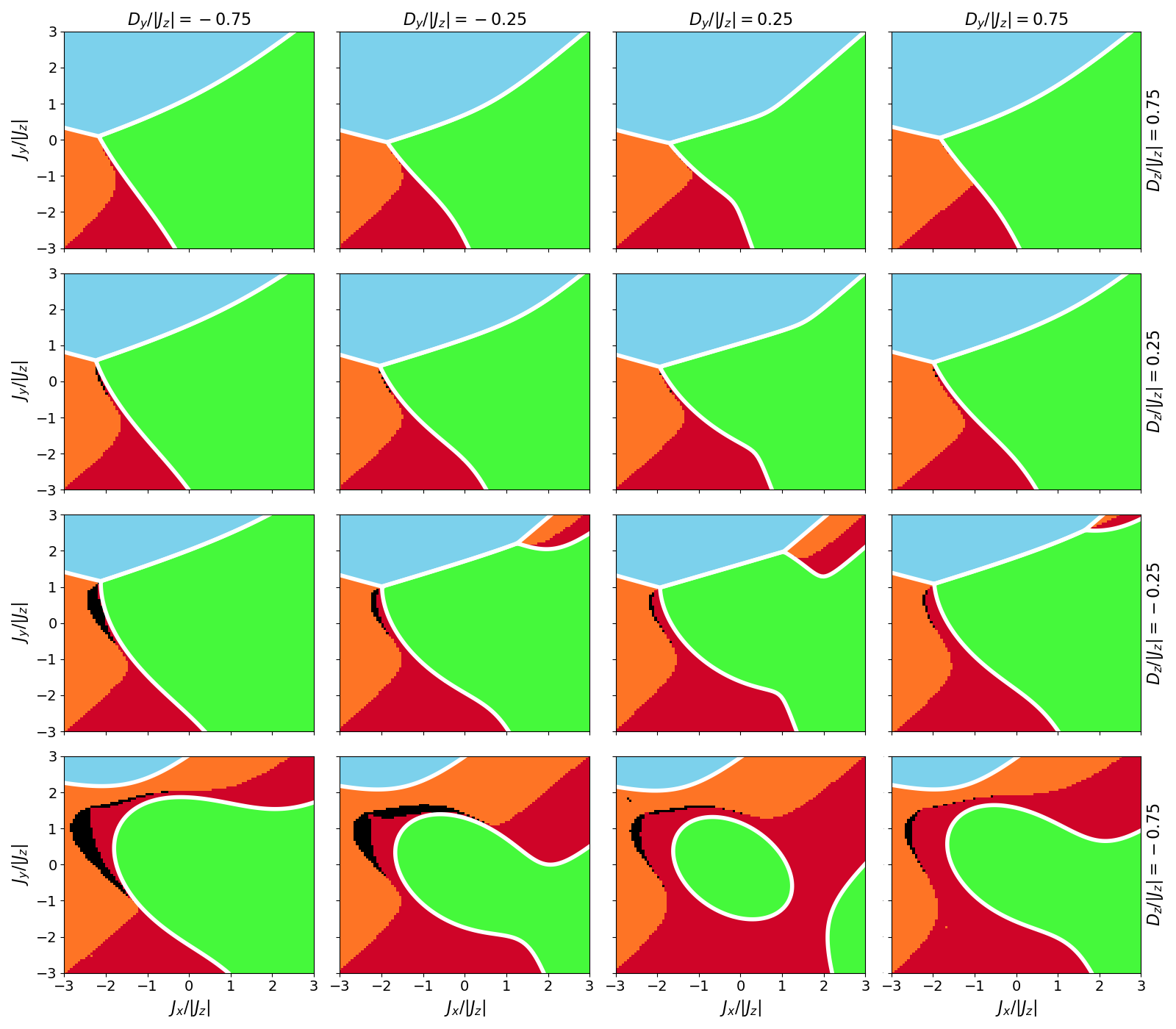}
\caption{$T=0$ phase diagram with $J_z<0$ and $K=-0.25 |J_z|$.
Each panel shows a slice of the phase diagram as a function of $J_x$ and $J_y$ for different, 
fixed, values of the DM directions $D_y$ and $D_z$, with $D_y$ increasing from left to right and
$D_z$ from bottom to top.
The phase diagram is obtained by comparing numerically optimized energies for the five phases described
in the main text.
The white lines show analytic calculations of the boundaries of the $\sf A_1$ and $\sf A_2$
phases.
}
\label{fig:pd_j-_k-025}
\end{figure*}

\begin{figure*}
\centering
\includegraphics[width=0.5\textwidth]{color_key.pdf}\\
\includegraphics[width=0.8\textwidth]{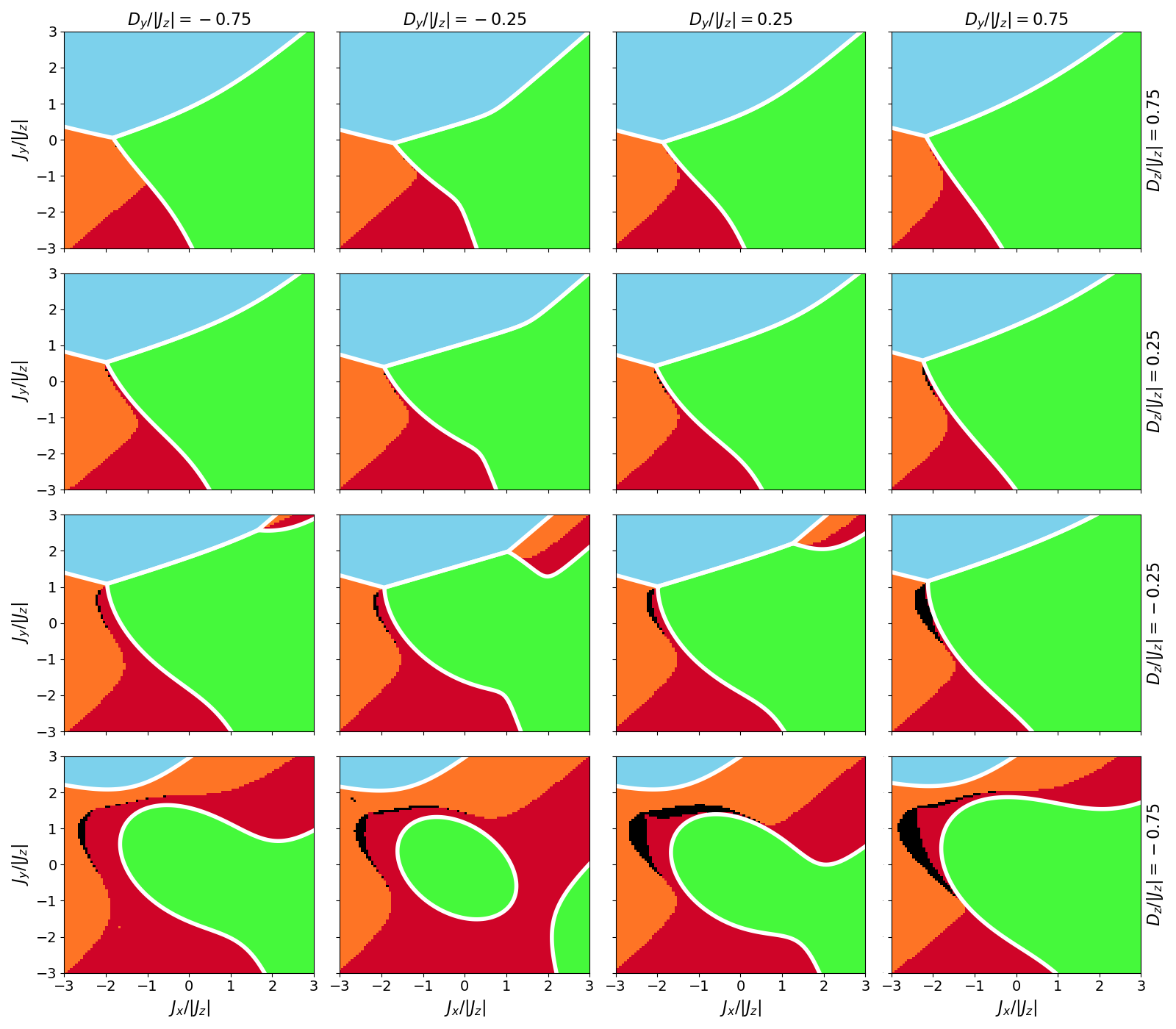}
\caption{$T=0$ phase diagram with $J_z<0$ and $K=0.25 |J_z|$.
Each panel shows a slice of the phase diagram as a function of $J_x$ and $J_y$ for different, 
fixed, values of the DM directions $D_y$ and $D_z$, with $D_y$ increasing from left to right and
$D_z$ from bottom to top.
The phase diagram is obtained by comparing numerically optimized energies for the five phases described
in the main text.
The white lines show analytic calculations of the boundaries of the $\sf A_1$ and $\sf A_2$
phases.
}
\label{fig:pd_j-_k+025}
\end{figure*}

\begin{figure*}
\centering
\includegraphics[width=0.5\textwidth]{color_key.pdf}\\
\includegraphics[width=0.8\textwidth]{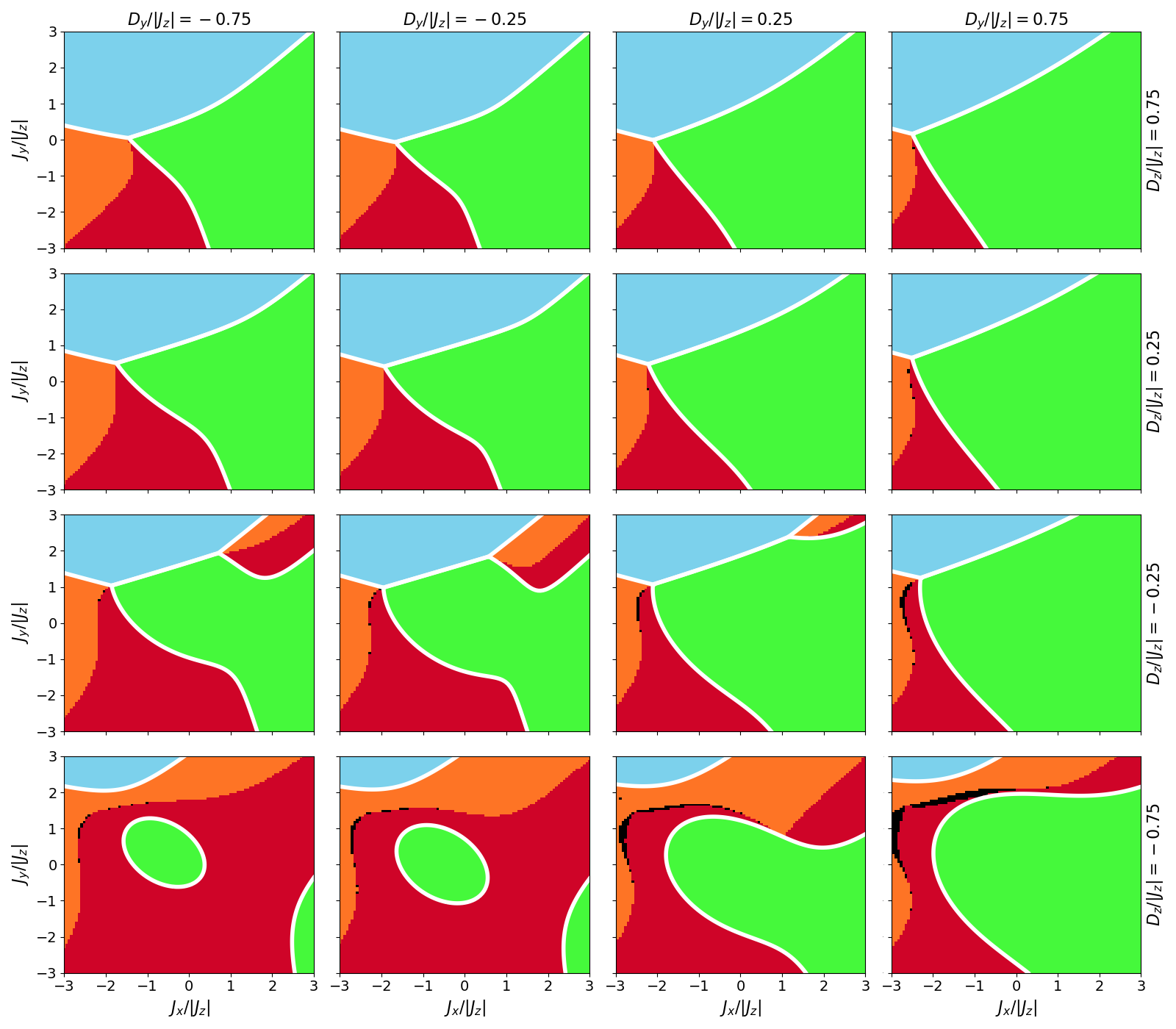}
\caption{$T=0$ phase diagram with $J_z<0$ and $K=0.75 |J_z|$.
Each panel shows a slice of the phase diagram as a function of $J_x$ and $J_y$ for different, 
fixed, values of the DM directions $D_y$ and $D_z$, with $D_y$ increasing from left to right and
$D_z$ from bottom to top.
The phase diagram is obtained by comparing numerically optimized energies for the five phases described
in the main text.
The white lines show analytic calculations of the boundaries of the $\sf A_1$ and $\sf A_2$
phases.
}
\label{fig:pd_j-_k+075}
\end{figure*}
